\documentclass[%
 reprint,amsmath,amssymb,
 aps,
 showkeys,
 onecolumn,
]{revtex4-2}
\usepackage{graphicx}
\usepackage{eqnarray} 
\usepackage{graphicx}
\usepackage{subcaption}
\usepackage{dcolumn}
\usepackage[normalem]{ulem}
\usepackage{cancel}
\usepackage{bm}
\usepackage{appendix}
\usepackage{xcolor}

\begin{document}

%-----------------------------------------------------------------------------
% Macros
%-----------------------------------------------------------------------------
% Abbreviations
\newcommand{\ie}{\mbox{\textit{i.e.}}}
\newcommand{\eg}{\mbox{\textit{e.g.}}}
\newcommand{\cf}{\textit{cf.}}

% Other
\newcommand{\bs}{\boldsymbol}
\newcommand{\dd}{\mathrm{d}}
\newcommand{\xh}{\hat{x}}
\newcommand{\yh}{\hat{y}}
\newcommand{\zh}{\hat{z}}

% For multiletter symbols
\newcommand\Rey{\mbox{\textit{Re}}} % Reynolds number
\newcommand\Rem{\mbox{\textit{Re}\textsubscript{m}}} % Magnetic Reynolds number
\newcommand\Prm{\mbox{\textit{Pr}\textsubscript{m}}} % Magnetic Prandtl number
\newcommand\Ha{\mbox{\textit{Ha}}} % Hartmann number
\newcommand\Nc{\mbox{$\mathcal{N}$}} % Stuart number
\newcommand\Ecf{E^{\mathrm{cf}}} % Crossflow-energy
\newcommand\Estk{E^{\mathrm{stk}}} % Streak-energy

% Vector quantities
\newcommand{\vv}{\bs v} % Velocity
\newcommand{\xv}{\bs x} % Coordinates
\newcommand{\jv}{\bs j} % Current density
\newcommand{\ee}{\bs e} % Unit vector
%-----------------------------------------------------------------------------

\title{The route to turbulence in magnetohydrodynamic square duct flow}
\thanks{Article accepted for publication in Phys.~Rev.~Fluids 18 December 2024,\\ \url{https://journals.aps.org/prfluids/accepted/1e07cSb8Hc017307a32578c36d10f348c95070931}.}

\author{Mattias Brynjell-Rahkola}
\email{Present address: Department of Applied Mathematics and Theoretical Physics, University of Cambridge, Centre for Mathematical Sciences, Wilberforce Road, Cambridge CB3 0WA, United Kingdom.\\
Corresponding author: mattias.brynjell-rahkola@tu-ilmenau.de}
\affiliation{%
Institut f{\"u}r Thermo- und Fluiddynamik, Technische Universit{\"a}t Ilmenau, Postfach 100565 D-98684 Ilmenau, Germany
}%
\author{Yohann Duguet}%
\affiliation{%
Laboratoire Interdisciplinaire des Sciences du Num\'erique -- LISN-CNRS, Universit\'e Paris-Saclay, F-91400 Orsay, France
}%

\author{Thomas Boeck}
\affiliation{
Institut f{\"u}r Thermo- und Fluiddynamik, Technische Universit{\"a}t Ilmenau, Postfach 100565 D-98684 Ilmenau, Germany
}%

\date{\today}

\begin{abstract}
The transition route from laminar to turbulent flow in a magnetohydrodynamic (MHD) duct with a square cross-section is investigated in the limit of low magnetic Reynolds number. In the presence of a transverse magnetic field, Hartmann and Shercliff layers are present on the walls orthogonal and parallel to the field direction, respectively.
We assume reflection symmetries in both transverse directions, and investigate the competition between transition mechanisms specific to each boundary layer using direct numerical simulations.
Independently of which wall turbulence eventually occupies, transition relies exclusively on a tripping of the Shercliff layer by perturbations, while the Hartmann layer plays a passive role. 
This is explained, using a dynamical systems interpretation, by the spatial localization of the edge states in the Shercliff layer at the expense of the Hartmann layer. 
The link between these non-linear coherent structures and the linear optimal modes known from non-modal stability and energy stability theory is pointed out.
\end{abstract}


\maketitle

%=============================================================================
\section{Introduction}

The physical mechanisms at play during the transition from laminar to turbulent shear flow have been a subject of investigation
for many decades. In many common circumstances, the situation is made difficult by the dynamical competition between several transition routes characterized by different flow physics. This is the case for instance in channel flow or in boundary layer flows with the competition between, on one hand, classical transition initiated by a modal instability scenario and, on the other hand, bypass transition which relies on non-modal instability mechanisms \cite{morkovin1969many}. This competition involves in its initial stages fully different coherent structures, respectively spanwise and streamwise vortices although the eventual turbulent flow is independent of the taken path. In the present study we are interested in another case of competition involving two non-modal scenarios.
We consider the incompressible flow inside an infinitely long square duct geometry. In standard hydrodynamic (HD) conditions, the associated laminar base flow is known to be linearly stable at all Reynolds numbers ($\Rey$) \cite{tatsumi1990stability,theofilis2004viscous}. Therefore transition to turbulence can only be triggered in the presence of finite-amplitude perturbations. All walls are equivalent by symmetry and the transition mechanisms do not depend on which wall that is disturbed initially.
This symmetric configuration is lost if the flow is, in addition, subject to an anisotropic force. One of the simplest examples of a force promoting anisotropy is the Lorentz force. We focus here on the case where the fluid is electrically conducting and the flow is subject to a homogeneous magnetic field imposed in a direction transverse to the mean flow direction and parallel to one pair of the sidewalls. This magnetohydrodynamic (MHD) configuration is common in many industrial magnetic flows of liquid metal. The associated Lorentz force brings anisotropy by introducing a difference in thickness and structure of the boundary layers at each wall depending on their orientation with respect to the magnetic field. The two walls orthogonal to the magnetic field are referred to as Hartmann walls \cite{hartmann1937hg}, and those parallel to it as Shercliff walls \cite{shercliff1953steady}. Their respective thicknesses in the laminar context scale, for large enough $\Ha$, like $O(\Ha^{-1})$ and $O(\Ha^{-1/2})$, respectively, where $\Ha$ is the Hartmann number proportional to the magnetic field intensity \cite{knaepen_moreau_2008}. The competition between these two boundary layers corresponds to the problem of understanding on which wall turbulence arises first as time increases. This knowledge is crucial in order to foster faster transition, or  to globally delay transition using \eg~passive control techniques. 

For the plane Hartmann channel, the laminar friction factor is given by 
\begin{equation}
    \label{eq:friction_factor}%
    \lambda = \frac{2\Ha^2}{\Rey}\left(\frac{\mathrm{tanh}(\Ha)}{\Ha-\mathrm{tanh}(\Ha)}\right)\sim \frac{2\Ha}{\Rey},
\end{equation}
for large values of $\Ha$ \cite{hartmann1937hg,branover_1978}, where $\Rey$ stands for the Reynolds number based on the bulk velocity. 
Measurements of friction factors for turbulent flow in large aspect ratio ducts (as a proxy for Hartmann channels) were reviewed in Ref.~\cite{branover_1978}. For sufficiently high Hartmann numbers, they were found to match the regime defined by \eqref{eq:friction_factor}, which is linear in $\Ha$. This was interpreted as a laminarization of the flow, which allowed to define the critical Reynolds number for laminar-turbulent transition, $\Rey_{\text{c}}=C\Ha$ with $C=2/\lambda_{\text{c}}$ as the point of departure from the linear relation \eqref{eq:friction_factor}.
While this is of little surprise in wide ducts it was reported to hold also for circular pipes and small aspect ratio ducts \cite{branover_1978}, which \emph{a priori} would suggest, also for these geometries, a transition scenario involving only the Hartmann layer.
Such a conclusion is at odds with more recent results, which suggest that long modes with low streamwise wavenumber $k_x$ localized within the Shercliff layers could be equally relevant \cite{krasnov_rossi_zikanov_boeck_2008}. 
Studies on transition to turbulence in isolated Hartmann layers have established that modal scenarios are unlikely at moderate Reynolds numbers \cite{lingwood1999stability,krasnov2004numerical}.
Instead, the transition is mediated, from a linearized point of view, by the transient growth of non-modal flow structures such as streamwise vortices and streamwise streaks. How these independent results combine together in the case of a duct geometry is currently not well understood. In the HD case, the linear non-modal approach has been applied to the square duct flow problem \cite{biau2008transition} and to corner flows \cite{schmidt2015optimal} with qualitatively similar results. 
This approach in the presence of MHD effects, has also been used in duct geometries \cite{krasnov2010optimal,dong2017linear,cassells2019three} and pipes \cite{aakerstedt1995damping,velizhanina2024optimal}. 
It indicates that linearly optimal modes, those maximizing the transient energy growth over a finite-time interval, are predominantly localized within the Shercliff/side layer. 
Recently, energy stability was carried out to identify the non-modal structures at the onset of transient growth, again pointing towards localization inside the Shercliff layers \cite{boeck_etal_2024}.
All these results support the prominent role played by the Shercliff layer, yet they remain based on linear theories.

We now wish to go beyond these suggestive results by investigating the transition problem from the \emph{nonlinear} point of view, common in modern HD transition studies. 
The nonlinear approach to transition for linearly stable flows started in the 1990s. The search for alternative (unstable) nonlinear solutions to the full governing equations, aimed at explaining why the flow could stay away from the laminar state in some well-defined state space \cite{kawahara2012significance,eckhardt2018transition}. In the HD duct flow case, several nonlinear traveling wave solutions were identified numerically using homotopy techniques \cite{wedin2008coherent,wedin2009three,uhlmann2010traveling,okino2010new,okino2012asymmetric}. At about the same time, it was realized that some unstable nonlinear solutions, namely edge states (see Ref.~\cite{biau2009optimal} in square duct), played a specific role in the transition process by mediating between the state space trajectories belonging to the laminar attraction basin and those leading to the turbulent state \cite{khapko2016edge}. Edge state computations have been performed first in channel \cite{itano2001dynamics} and pipe flow \cite{schneider2007turbulence,eckhardt2008dynamical,duguet2008transition} followed by the square duct case \cite{biau2009optimal}. 
Their generalization to MHD configurations is straightforward from a theoretical point of view but remain scarce in the literature \cite{guseva2015transition,shuai2023exact,li2024effects,brynjell-rahkola_etal_2024}.
The identification of edge states is a first step towards a rigorous identification of effective transitioning trajectories associated with the unstable manifold of the edge state solutions \cite{duguet2010slug}. Once such transitioning trajectories are known, they constitute an ideal laboratory to untangle the transition mechanisms, without relying on any linearization assumption \cite{duguet2013minimal}. 
One crucial property of all edge state solutions reported so far in HD is their spatial localization, which emerges as soon as the computational domain is large enough \cite{duguet2009localized,schneider2010localized}. 
This property makes edge states physically robust objects independent of the numerical domain used to simulate them. Besides, the active region of these solutions can be interpreted as a specific region where early transition to turbulence is favored, the transition at other locations following in time through a spatial contamination process \cite{duguet2013minimal}. In this study we  exploit this localization property precisely to discuss on which wall of the MHD duct geometry transition happens and why. 

In order to lower the complexity of the transition phenomenon as well as the computational cost, the spatial symmetries inherent to the square duct geometry are considered and imposed in the computations. An outline of this \emph{symmetric model} along with flow parameters and numerical method considered is given in \S\ref{sec:method_model}. We follow the nonlinear road map by computing and describing active transition scenarios and the edge states in \S\ref{sec:transition}. 
The results are finally summarized in \S\ref{sec:conclusions_outlook}.

%=============================================================================
\section{Flow model and numerical method}
\label{sec:method_model}%

\begin{figure}
    \includegraphics[width=0.35\textwidth]{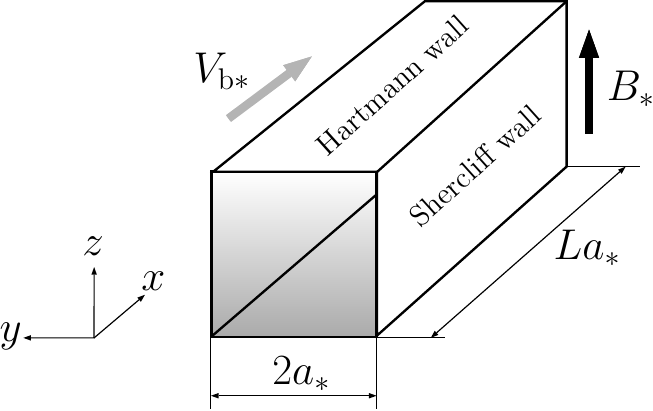}
    \caption{Sketch of the numerical set-up with the defining quantities used in the non-dimensionalization.}
    \label{fig:sketch_setup}%
\end{figure}

\subsection{Governing equations}
\label{sec:goveq}%

We consider the flow of an electrically conducting fluid inside a straight square duct (unit aspect ratio) with electrically insulating walls (see Fig.~\ref{fig:sketch_setup}). 
The coordinate system has the streamwise direction $x$ and the two wall-normal directions $y$ and $z$. The flow is subject to a magnetic field $B_*\ee_z$ along the $z$-axis, which is labeled as the \emph{vertical} direction. (The subscript '$*$' denotes dimensional quantities).
Hartmann and Shercliff boundary layers are present along the walls orthogonal to the $z$- and $y$-axis, respectively, as shown in Fig.~\ref{fig:bf_vx} for a laminar velocity field. 
Throughout this article, we will refer to these walls simply as \emph{Hartmann} and \emph{Shercliff walls} as indicated in Fig.~\ref{fig:sketch_setup}, and denote physical quantities evaluated on these walls with a superscript 'H' and 'S', respectively.

The flow is governed by the incompressible Navier-Stokes equations subject to the Lorentz force in the low-$Re_m$ approximation, in which the induction equation for the magnetic field need not be considered. The duct half-width $a_*$ and the bulk velocity $V_{\mathrm{b}*}$ (\ie~the streamwise velocity averaged over a cross-section), which are constant in time, are used for non-dimensionalization. 
This defines the advective time scale $a_*/V_{\mathrm{b}*}$.
The two independent non-dimensional parameters of the problem are the bulk Reynolds number $\Rey=V_{\mathrm{b}*}a_*/\nu_*$ and the Hartmann number $\Ha=B_* a_*(\mu_*\eta_*\rho_*\nu_*)^{-1/2}$, where $\nu_*$, $\eta_*$, $\rho_*$ and $\mu_*$ in turn are the fluid's kinematic viscosity, its magnetic diffusivity, its density, and the magnetic permeability of free space.

These parameters form the magnetic interaction parameter $\Nc=\Ha^2/\Rey$, also called the Stuart number. 
It is the ratio between electromagnetic and inertial forces, or alternatively between the advective time scale $a_*/V_{\mathrm{b}*}$ and the magnetic damping time scale $\rho_*\mu_*\eta_*/B_*^2$ \cite{davidson_2016}.
In reference to liquid metals, the magnetic Prandtl number $\Prm=\nu_*/\eta_*$, which represents the ratio between the kinematic viscosity and the magnetic diffusivity, is small. 
Assuming a moderate ($O(10^4)$ or smaller) value for $\Rey$, the magnetic Reynolds number $\Rem=\Rey\Prm$ is small as well.
This is the case in most laboratory and industrial flows \cite{knaepen_moreau_2008} and allows for the magnetic self-induction to be neglected.
Such a framework is commonly referred to as the \emph{quasi-static MHD approximation}, in which the magnetic field is taken to be steady and irrotational.
Under these choices, it can be shown (see \eg~\cite{boeck_2010}) that the Lorentz force is approximated at first order through Ohm's law with the electric field expressed as the gradient of a scalar potential, and a charge conservation condition.
The non-dimensional governing equations read
\begin{subequations}
    \label{eq:governing_eqs}%
    \begin{align}
        \label{eq:navierstokes}%
        \frac{\partial \vv}{\partial t} + (\vv\cdot\nabla)\vv &= -\nabla p + \frac{1}{\Rey}\nabla^2\vv
        + \Nc(\jv\times\ee_z) + \chi\ee_x,\\
        \label{eq:ohmslaw}%
        \jv &= -\nabla\phi + \vv\times\ee_z,\\
        \label{eq:mass_conservation}%
        \nabla\cdot\vv &= 0,\\
        \label{eq:charge_conservation}%
        \nabla\cdot\jv &= 0,
    \end{align}
\end{subequations}    
where $\vv$ is the velocity, $\jv$ is the the current density, $p$ is the pressure and $\phi$ is the electric potential. The additive term $\chi$ is a spatially uniform time-dependent volume force applied in the streamwise direction $x$ to enforce a constant mass flux. 
All walls obey the no-slip condition ${\bm v}|_{y=\pm 1}={\bm v}|_{z=\pm 1}={\bm 0}$, and are electrically insulating, \ie~\mbox{$(\jv\cdot\ee_y)|_{y=\pm 1}=(\jv\cdot\ee_z)|_{z=\pm 1}=0$}.
Streamwise periodicity is assumed with an imposed wavelength $L$. 
The choice of the values of $\Rey$, $\Nc$ and $L$ is discussed in \S\ref{sec:parameter_values}.

\subsection{Base state}

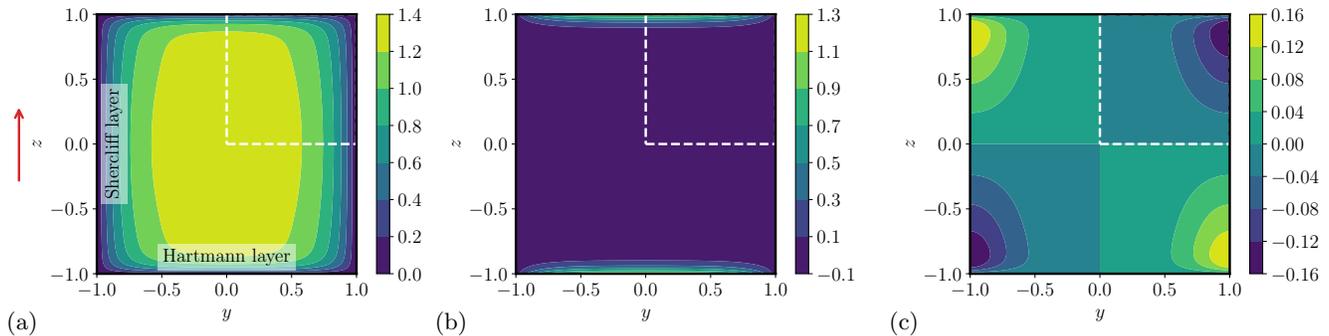
\begin{figure}
    \centering
      \begin{subfigure}[b]{0.31\textwidth}
        \setlength{\unitlength}{1.0cm}
        \centering
        \begin{picture}(5.15,4.3)
            \put(-0.15,0){\includegraphics[scale=0.45]{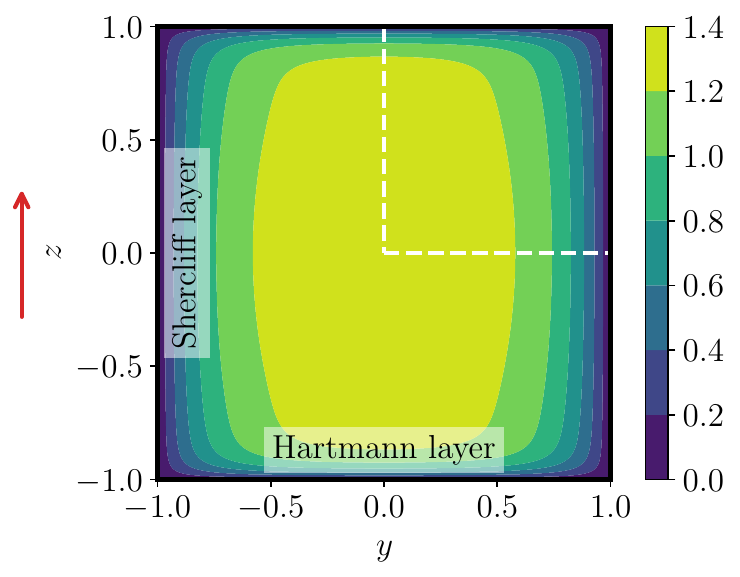}}%
            \put(-0.15,0.0){(a)}
        \end{picture}       
        \phantomsubcaption
        \label{fig:bf_vx}%
      \end{subfigure}
      \begin{subfigure}[b]{0.33\textwidth}
        \setlength{\unitlength}{1.0cm}      
        \centering
        \begin{picture}(5.45,4.3)
            \put(-0.1,0){\includegraphics[scale=0.45]{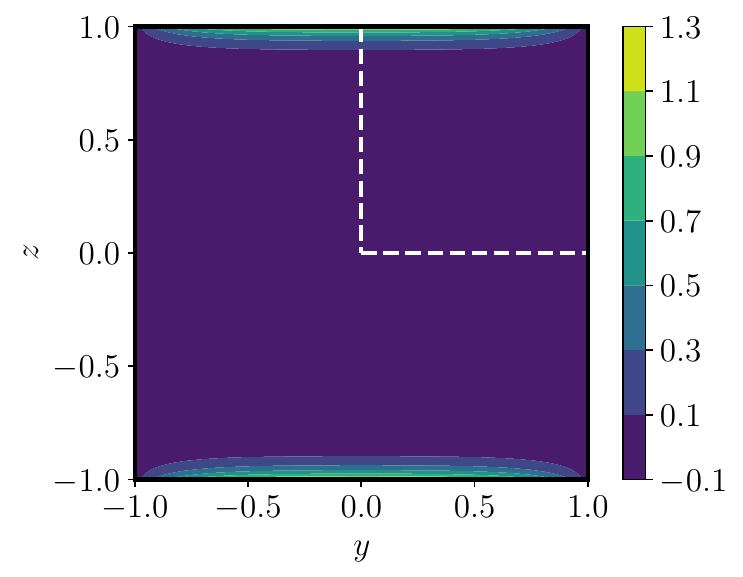}}%
            \put(-0.15,0.0){(b)}
        \end{picture}
        \phantomsubcaption
        \label{fig:bf_jy}%
      \end{subfigure}
      \begin{subfigure}[b]{0.34\textwidth}
        \setlength{\unitlength}{1.0cm}         
        \centering
        \begin{picture}(5.6,4.3)
            \put(-0.1,0){\includegraphics[scale=0.45]{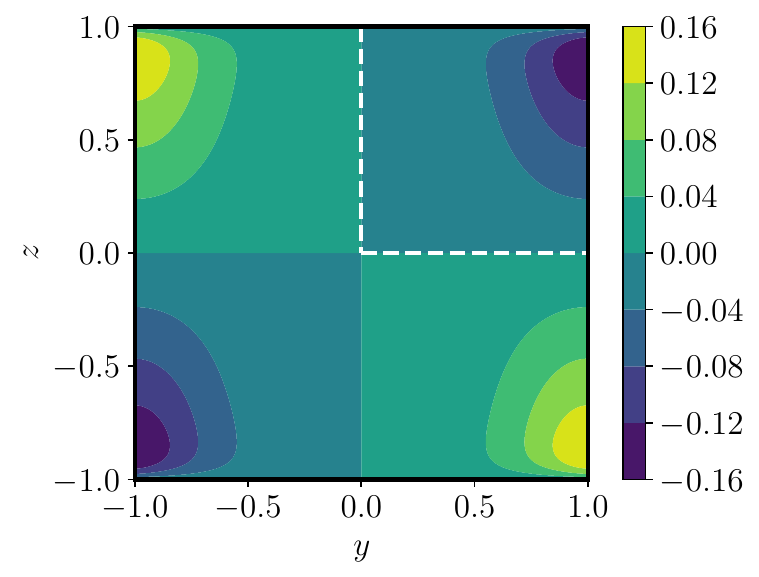}}%
            \put(-0.15,0.0){(c)}
        \end{picture} 
        \phantomsubcaption
        \label{fig:bf_jz}%
      \end{subfigure}
      \caption{
      Laminar base state.
      Analytical solution to \eqref{eq:baseflow_eq} for $\Ha=20$ \cite{mueller_buehler_2002} scaled to match the non-dimensionalization adopted in the present work.
      Isocontours of (\subref{fig:bf_vx}) the streamwise velocity $\widetilde{v}_x$,
      (\subref{fig:bf_jy}) the transverse current density $\widetilde{j}_y$,
      (\subref{fig:bf_jz}) the current density component aligned with the magnetic field $\widetilde{j}_z$.
      The direction of the magnetic field is shown with a red arrow in (\subref{fig:bf_vx}) and the white dashed square marks the fundamental domain in the symmetry reduction of \S\ref{sec:symred_model}}.
      \label{fig:baseflow}%
\end{figure}

The steady laminar flow can be obtained analytically without the assumption of low magnetic Reynolds numbers. It is computed using the full induction equation and involves the streamwise velocity $\widetilde{v}_x$ and the streamwise induced magnetic field $\widetilde{b}_x$.
It satisfies the following set of equations
\begin{subequations}
  \label{eq:baseflow_eq}%
  \begin{gather}
    \frac{\partial^2 \widetilde{v}_x}{\partial y^2} + \frac{\partial^2 \widetilde{v}_x}{\partial z^2} + \Ha\frac{\partial \widetilde{b}_x}{\partial z} = -1, \qquad
    \frac{\partial^2 \widetilde{b}_x}{\partial y^2} + \frac{\partial^2 \widetilde{b}_x}{\partial z^2} + \Ha\frac{\partial \widetilde{v}_x}{\partial z} = 0.
    \tag{\theequation a,b}
  \end{gather}
\end{subequations}
Throughout the paper, quantities related to the base flow are denoted with a tilde, $\widetilde{(\cdot)}$.
The solution to \eqref{eq:baseflow_eq} was originally found by Shercliff \cite{shercliff1953steady} in the form of infinite series, and later re-derived in \eg~\cite{mueller_buehler_2002}.
The convergence of these series was recently discussed in \cite{brynjell-rahkola_2024}.
Note that the non-dimensionalization in eq.~\eqref{eq:baseflow_eq} is based on the prescribed pressure gradient and therefore different from that introduced in \S\ref{sec:goveq} (see Ref.~\cite{mueller_buehler_2002} for details).
From the expression for the induced flux density, the current density components are obtained as
\begin{subequations}
    \label{eq:jy_jz}%
    \begin{gather}
      \widetilde{j}_y = \frac{1}{\Ha}\frac{\partial \widetilde{b}_x}{\partial z}, \qquad
      \widetilde{j}_z = -\frac{1}{\Ha}\frac{\partial \widetilde{b}_x}{\partial y}.   
      \tag{\theequation a,b}
    \end{gather}
\end{subequations}
The velocity field $\widetilde{v}_x$ is shown in Fig.~\ref{fig:baseflow} together with the current components $\widetilde{j}_y$ and $\widetilde{j}_z$ 
(all rescaled to match the non-dimensionalization introduced in \S\ref{sec:goveq}). 
The distinction between the narrower $O(\Ha^{-1})$ horizontal Hartmann layers near $z=\pm 1$ and the thicker $O(\Ha^{-1/2})$ vertical Shercliff layers near $y=\pm 1$ is attested in the figure. The transverse current component $\widetilde{j}_y$ is small and negative in the bulk region but rises to $O(1)$ in the Hartmann layers, where it accelerates the flow.

The corresponding laminar solution in the HD square duct is known to be linearly stable for all $\Rey$ \cite{tatsumi1990stability}.
In contrast, Hunt's flow, which corresponds to a square duct with electrically insulating and conducting walls parallel and orthogonal to the magnetic field, respectively,
is found to be linearly unstable above $\Ha\approx 5.7$ \cite{priede2010linear}.
The case of a rectangular duct with electrically insulating walls and aspect ratio 5 has been studied in \cite{tagawa2019linear}. However, the authors are not aware of any linear stability reported in the literature for the square MHD case.
Instead, this flow supports significant transient growth \cite{krasnov2010optimal}, and is believed to transition to a sustained turbulent state via a subcritical scenario involving finite-amplitude  perturbations.

\subsection{Numerical protocol}
\label{sec:num_sim}%

In the general unsteady case, equations \eqref{eq:governing_eqs} are tackled with direct numerical simulation (DNS) using the open source code \textit{NEK5000} \cite{nek5000}, known for its high accuracy and parallel performance.
Specifically, equations \eqref{eq:governing_eqs} are discretized using the staggered $\mathbb{P}_N$-$\mathbb{P}_{N-2}$ spectral element method (SEM) \cite{maday_patera_1989}, where the domain is divided into non-overlapping hexahedral elements, and the solution variables are expanded in high-order Lagrange interpolation polynomials.
The velocity and pressure are represented on $N+1$ Gauss-Lobatto-Legendre, and $N-1$ Gauss-Legendre quadrature nodes, respectively.
The $\mathbb{P}_N$-$\mathbb{P}_{N-2}$ scheme was recently extended to the $\jv$-$\phi$ quasi-static MHD formulation in \cite{brynjell-rahkola_2024}, where the exponential convergence of the method was verified. The polynomial order $N$ is here chosen to be either 7 or 11 (see Appendix \ref{sec:resolutions} for details).
Evaluation of the nonlinear advection terms is done using over-integration \cite{malm2013stabilization}. As described in \cite{brynjell-rahkola_2024}, the equations are integrated in time using a third order backwards-differentiation/extrapolation scheme \cite{karniadakis_israeli_orszag_1991}.

\subsection{Symmetric model}
\label{sec:symred_model}%

Many recent transition studies have exploited the possibility to impose discrete symmetries to the system in order to simplify both the temporal dynamics and the computational load. 
Such an approach has for instance been successful in uncovering nonlinear solutions in HD pipe \cite{duguet2008transition,avila2013streamwise} and duct flow \cite{uhlmann2010traveling}.
This is based on the idea that the stability of the relevant symmetric solutions is likely to be increased by the restriction of the dynamics to a well-chosen subspace. In this section we suggest a symmetric model of the MHD duct flow obtained by imposing two discrete symmetries at the same time. Continuous symmetries, here the equivariance of the governing equations and boundary conditions with respect to streamwise translations, have not been 
considered. We define
\begin{subequations}
    \label{eq:symrel_vel}%
    \begin{align}
        \label{eq:symrel_1y}%
        s_{1y}:\,[v_x, v_y, v_z, p](y,z) &=
        [v_x, -v_y, v_z, p](-y,z), \\
        \label{eq:symrel_z1}%
        s_{1z}:\,[v_x, v_y, v_z, p](y,z) &=
        [v_x, v_y, -v_z, p](y,-z).
    \end{align}
\end{subequations}
As visible in Fig.~\ref{fig:bf_vx}, the laminar base flow $\widetilde{v}_x$ satisfies reflection symmetry conditions with respect to the $y$- and $z$-axis.
The symmetries of the current density components depend on the direction of the magnetic field.
In the present case where the magnetic field is aligned with the $z$-axis, $\widetilde{\jv}$ satisfies a simple reflection symmetry with respect to the $z$-coordinate and a reflection symmetry followed by a sign flip with respect to the $y$-coordinate,
\begin{subequations}
    \label{eq:symrel_cur}%
    \begin{align}
        \label{eq:symrel_y2}%
        s_{2y}:\,[j_x, j_y, j_z, \phi](y,z) &=
        [-j_x, j_y, -j_z, \phi](-y,z), \\
        \label{eq:symrel_z2}%
        s_{1z}:\,[j_x, j_y, j_z, \phi](y,z) &=
        [j_x, j_y, -j_z, \phi](y,-z).
    \end{align}
\end{subequations}
We note that these symmetries are also shared with the turbulent mean flows obtained in the numerical simulations in Refs.~\cite{huser1993direct, uhlmann2010traveling,pinelli2010reynolds,krasnov2012numerical,vinuesa2014aspect}. 
These symmetries are compatible with Prandtl's secondary flows  of the second kind \cite{bradshaw1987turbulent} emanating from the corner regions. This motivates the consideration of such symmetries both in order to reduce the cost of the simulations and to simplify the complexity of the transition process.
In practice, since the governing equations in the SEM are formulated and solved in weak form, imposing the above conditions in the numerical simulations amounts to a suitable change in the test and trial functions (see \eg~\cite{deville_fischer_mund_2002} for details).
Moreover, although no explicit boundary or symmetry conditions are imposed on the pressure and the potential field, a consistent solution will implicitly satisfy the relations \eqref{eq:symrel_vel} and \eqref{eq:symrel_cur}.

In the absence of any magnetic field, eigenfunctions of the classical hydrodynamic stability problem in a square duct flow have been found to belong to one of four independent symmetries denoted I-IV \cite{tatsumi1990stability}.
These symmetry classes read 
\begin{equation}
    \label{eq:sym_classes}%
    \begin{array}{rcccc}
        \text{I:}   & v_x(\mathrm{o,e}) & v_y(\mathrm{e,e}) & v_z(\mathrm{o,o}) & p(\mathrm{o,e}) \\
        \text{II:}  & v_x(\mathrm{o,o}) & v_y(\mathrm{e,o}) & v_z(\mathrm{o,e}) & p(\mathrm{o,o}) \\
        \text{III:} & v_x(\mathrm{e,e}) & v_y(\mathrm{o,e}) & v_z(\mathrm{e,o}) & p(\mathrm{e,e}) \\     
        \text{IV:}  & v_x(\mathrm{e,o}) & v_y(\mathrm{o,o}) & v_z(\mathrm{e,e}) & p(\mathrm{e,o})
    \end{array}
\end{equation}
where the first letter refers to the symmetry ('e' for even, 'o' for odd) with respect to the $y$-coordinate, and the second to the symmetry with respect to the $z$-coordinate \cite{uhlmann2006linear,uhlmann2010traveling}.
In terms of this nomenclature, the symmetry relations \eqref{eq:symrel_vel} and \eqref{eq:symrel_cur} imply that the function space becomes limited to the states for which the velocity and pressure belong to the symmetry class III, while the current and the potential belong to the symmetry class I. 
We will refer to this symmetric domain, for which $\{0 \le y,z \le 1\}$, as the \emph{quarter duct}. 
By contrast, the original system where no symmetry is imposed is referred to as the \emph{full duct}.
With the exception of Appendix \ref{sec:turb_stat}, we will consider solutions of \eqref{eq:governing_eqs} inside the finite domain $\Omega=[0,L)\times[0,1]\times[0,1]$.

\subsection{Parameter values}
\label{sec:parameter_values}%

\subsubsection{Hartmann and Reynolds numbers}
\label{sec:ha_re_numbers}%

Throughout this study, the Hartmann number is fixed to $\Ha=20$, which is large enough to yield distinct dynamics in the Shercliff and the Hartmann layers. 
In keeping $\Ha$ constant, the base flow shown in Fig.~\ref{fig:baseflow} remains unchanged. 
Therefore, the only way to vary the interaction parameter $\Nc$ and investigate the influence of a varying Lorentz force is to vary the bulk Reynolds number $\Rey$. 
The appearance of turbulence in MHD duct flow is commonly characterized by a Hartmann layer Reynolds number $R$, defined as the ratio between the bulk Reynolds and the Hartmann number, \ie~$R=\Rey/\Ha=V_{\mathrm{b}*}\delta_*^{\mathrm{H}}/\nu_*$, where $\delta_*^{\mathrm{H}} =\sqrt{\mu_*\eta_*\rho_*\nu_*}/B_*$ represents the Hartmann layer thickness  \cite{murgatroyd1953cxlii,zikanov2014laminar}.
From \cite{krasnov2013patterned}, it is known that long pipes and ducts with aspect ratio close to unity feature laminar Hartmann layers and turbulent/intermittent side/Shercliff layers for $R\approx 227$.
Furthermore, it was shown in \cite{moresco2004experimental,krasnov2004numerical} that turbulence may be sustained in the Hartmann channel at Reynolds numbers as low as $350\lesssim R\lesssim400$. 
Based on these findings, two combinations of parameter values are chosen: $(\Rey,\Ha)=(5000,20)$, $R=250$; and $(8000,20)$, $R=400$.
These parameter configurations correspond to $\mathcal{N}=0.08$ and $0.05$, respectively, which are far from the parameter regime $\mathcal{N}\gg 1$ where quasi-two-dimensional dynamics may be expected \cite{cassells2019three,camobreco2020subcritical}.

\begin{figure}
    \includegraphics[scale=0.5]{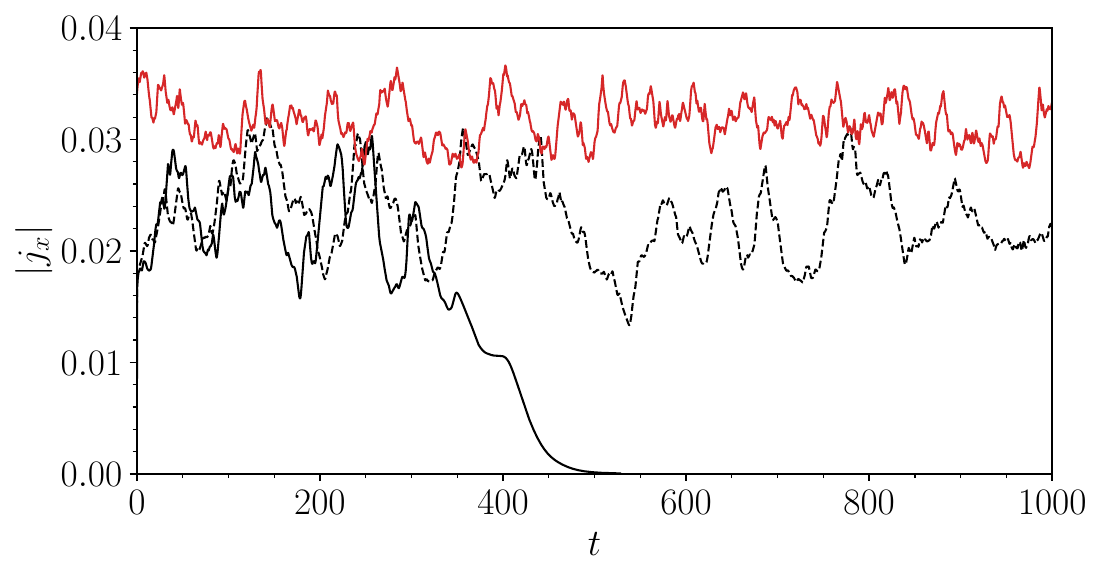}
    \caption{Sustenance of turbulence monitored by the streamwise current density $j_x$, for parameters $R=250$ (black) and $R=400$ (red) in domains with periodicity $L=2\pi$ (dashed) and $L=\pi$ (solid).}
    \label{fig:turb_monitor}%
\end{figure}

\subsubsection{Domain length}
\label{sec:length_domain}%

In addition to the symmetry relation \eqref{eq:symrel_vel} and \eqref{eq:symrel_cur}, the system of equations \eqref{eq:governing_eqs} is equivariant under the action of translations in the $x$-direction.
In order to isolate the active transition mechanisms, the concept of a minimum flow unit \cite{jimenez_moin_1991,hamilton1995regeneration} is particularly useful.
This is defined as a computational domain that is as small as possible in the planar (periodic) directions, yet large enough to support turbulence.
For the present duct geometry, it implies a reduction in the streamwise extent $L$.

To formalize the discussion, it is convenient to introduce two viscous length units specific to each wall, defined respectively by $x_+^{\mathrm{S}} = \Rey_{\tau}^{\mathrm{S}} x$ and $x_+^{\mathrm{H}}=\Rey_{\tau}^{\mathrm{H}}x $. Parameters $\Rey_{\tau}^{\mathrm{S}}$ and $\Rey_{\tau}^{\mathrm{H}}$ correspond to the friction Reynolds number on the Shercliff and the Hartmann walls, respectively. Following \cite{krasnov2012numerical}, we define them as
\begin{subequations}
  \label{eq:turbulence_quantities}%
  \begin{gather}
    \tau^{\mathrm{H}}(y,z) = \overline{\left\langle\frac{\partial v_x}{\partial z}\right\rangle_x}(y,z), \quad
    \tau_w^{\mathrm{H}}(y) = \left.\tau^{\mathrm{H}}(y,z)\right|_{y,z=1}, \quad
    u_{\tau}^{\mathrm{H}}(y) = \sqrt{\frac{\tau_w^{\mathrm{H}}(y)}{\Rey}}, \quad
    \Rey_{\tau}^{\mathrm{H}}(y) = \Rey\,u_{\tau}^{\mathrm{H}}(y)
    \tag{\theequation a,b,c,d}
  \end{gather}
\end{subequations}
on the Hartmann walls with analogous expressions for the Shercliff walls.
In \eqref{eq:turbulence_quantities}, $\tau$ is 
the dominant shear stress component of the mean flow,
$\tau_w$ is its value on the wall and $u_{\tau}$ is the friction velocity \cite{pope2001turbulent}. Note that $\langle\cdot\rangle_x$ refers to streamwise averaging (similar notation will be used to indicate averaging in other directions), and $\overline{(\cdot)}$ denotes time averaging.
Note also that $\Rey_{\tau}$ varies over the circumference of the duct due to the two inhomogeneous directions $y$ and $z$. 
An overview of this variation is given in Table \ref{tab:retau} of Appendix \ref{sec:turb_stat}.

The concept of minimal flow unit was investigated for the HD square duct in \cite{uhlmann2007marginally}, where it was shown that the critical streamwise period for sustaining turbulence is just below $L_+=200$.
The same idea to select a minimal domain was tested here in the MHD case $R=250$, by directly comparing the long-time dynamics for different values of $L=\{8\pi, 4\pi$, $2\pi$\}. 
The state of the flow is monitored in time using the normalized scalar $L^2$-norm of the streamwise current $j_x$, 
\begin{equation}
    \label{eq:jx_norm}%
    |j_x| = \sqrt{\frac{1}{L}\int_{\Omega} j_x^2\,\dd \upsilon}, 
\end{equation}
which according to Ohm's law \eqref{eq:ohmslaw} vanishes in the absence of secondary flow.
As shown in Fig.~\ref{fig:turb_monitor}, the turbulence may be sustained down to values of $L=2\pi$, which corresponds to $(L_+^{\mathrm{S}},L_+^{\mathrm{H}})=(1648,2058)$ using the wall-averaged values of $\Rey_{\tau}^{\mathrm{S}}$ and $\Rey_{\tau}^{\mathrm{H}}$ in Table \ref{tab:retau} (Appendix \ref{sec:turb_stat}).
Upon further reducing the periodicity to $L=\pi$, the flow laminarizes.
In terms of the current density, $|j_x|$ decays to $O(10^{-4})$ in around 500 time units (see \S\ref{sec:goveq}) and continues to decay beyond this.
Although the time required to reach this level changes slightly with variations in the initial conditions and the solver settings, the phenomenon is robust, which suggests that turbulence is simply not sustained at these parameter values. 
The minimal flow unit may thus be concluded to be considerably larger than in the HD case \cite{uhlmann2007marginally}.
Upon increasing the Reynolds number here to $\Rey=8000$ ($R=400$), turbulence appears sustained also in the $\pi$-periodic domain.
In this case the domain size is $(L_+^{\mathrm{S}},L_+^{\mathrm{H}})=(1368,1494)$.
The two parameters cases $(R,L)=(250,2\pi)$  and $(R,L)=(400,\pi)$ are summarized in Table \ref{tab:cases} respectively as cases A and B.

\begin{table}
    \caption{Summary of the investigated parameter configurations.}
    \label{tab:cases}%
    \begin{ruledtabular}
        \begin{tabular}{c c c c c c} 
            Case & $L$  & $\Rey$ & $\Ha$ & $R$ & $\mathcal{N}$ \\ 
            \colrule
            A & 2$\pi$ &  5000 & 20 & 250 & 0.08 \\ 
            B  & $\pi$ & 8000 & 20 & 400 & 0.05\\ 
        \end{tabular}
    \end{ruledtabular}
\end{table}

\begin{figure}
    \centering
    \begin{subfigure}[b]{0.52\textwidth}
        \setlength{\unitlength}{1.0cm}
        \centering
        \begin{picture}(8.6,7.0)
            \put(-0.2,0.0){\includegraphics[scale=0.5]{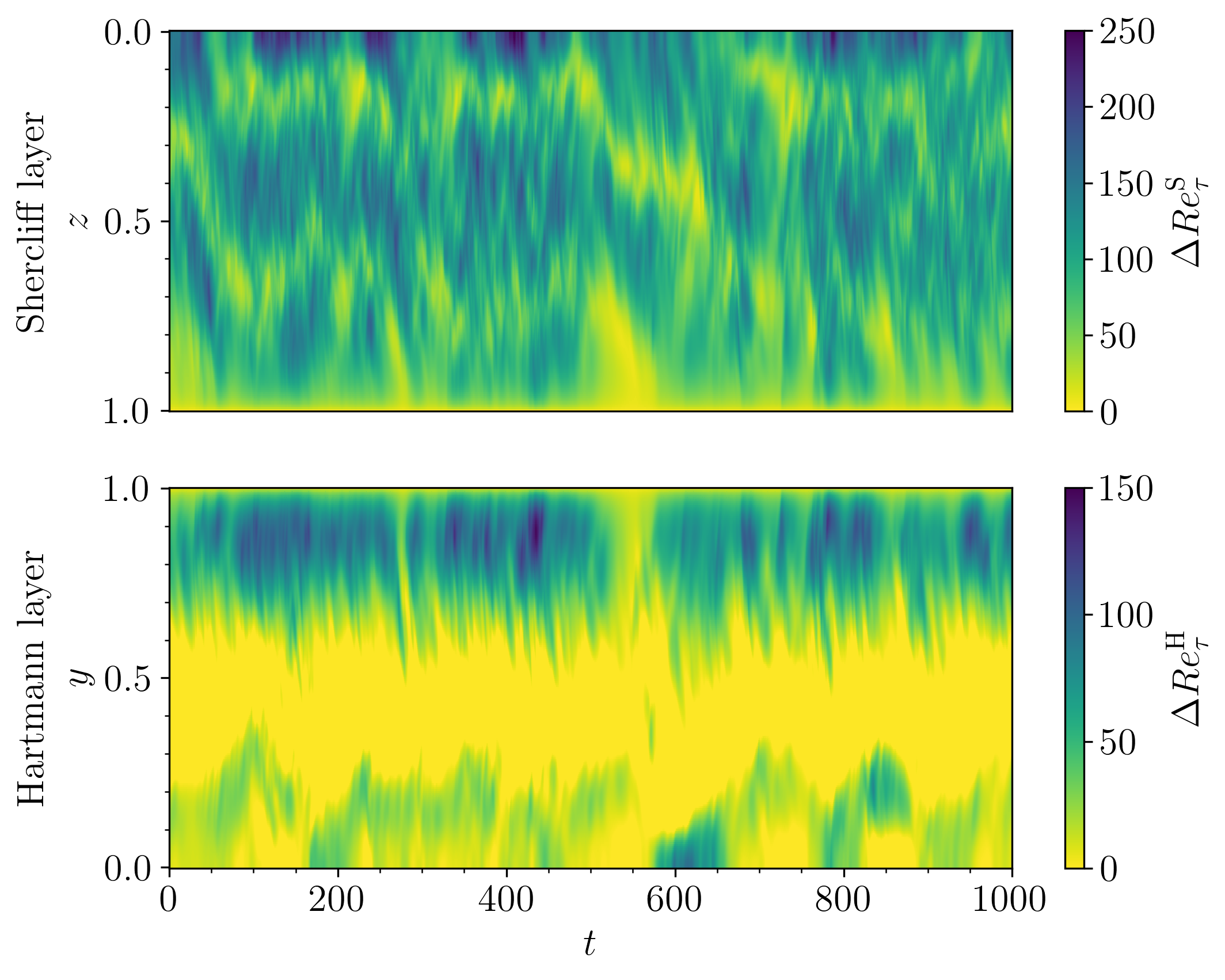}}%
            \put(-0.2,0.0){(a)}
        \end{picture}      
        \phantomsubcaption
        \label{fig:std_turb_re5k}%
    \end{subfigure}
    \begin{subfigure}[b]{0.47\textwidth}
        \setlength{\unitlength}{1.0cm}
        \centering
        \begin{picture}(7.7,7.0)
            \put(-0.1,0.0){\includegraphics[scale=0.5]{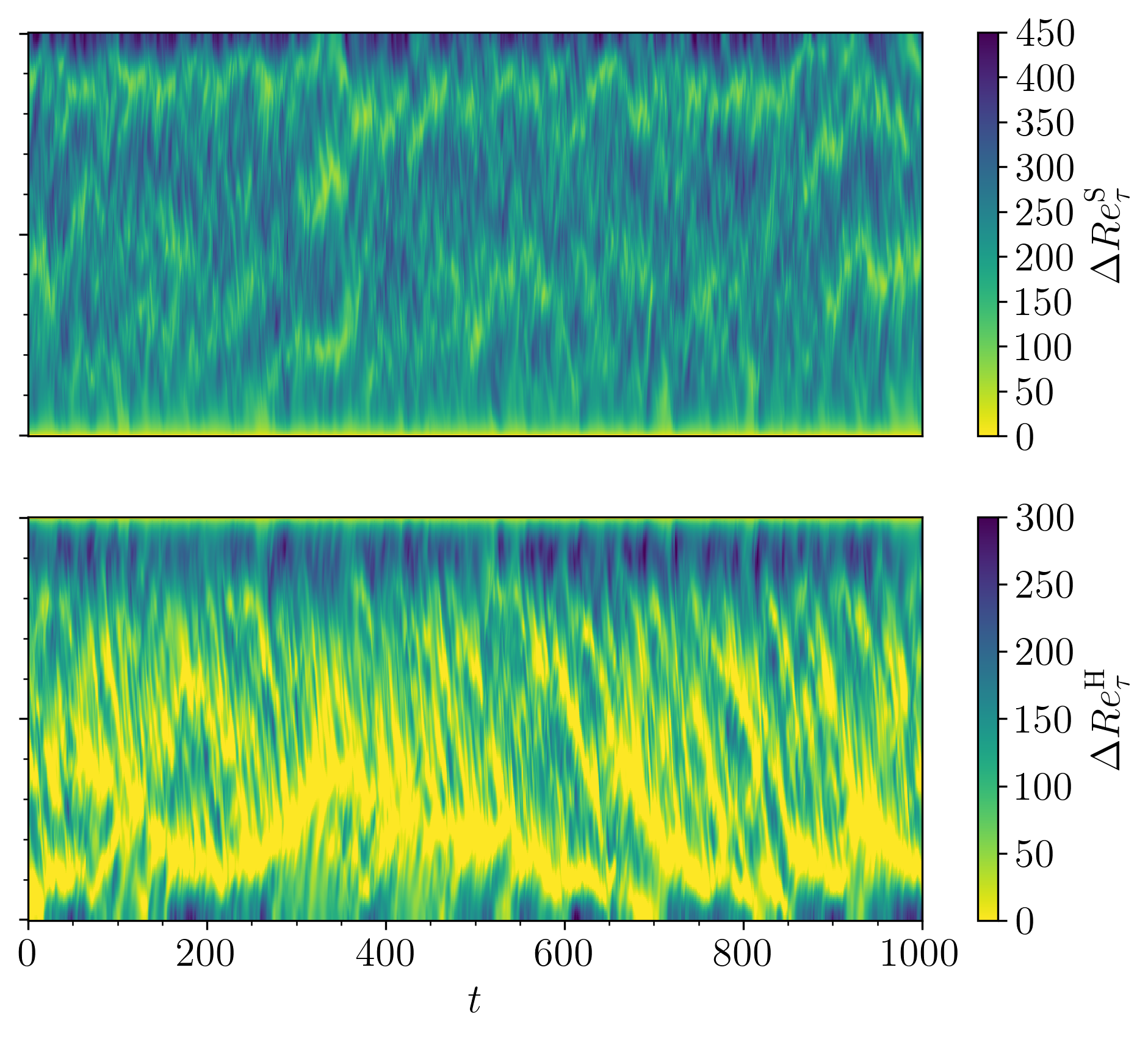}}%
            \put(-0.1,0.0){(b)}
        \end{picture}
        \phantomsubcaption
        \label{fig:std_turb_re8k}%
    \end{subfigure}    
    \caption{
    Space-time diagrams of excess friction Reynolds number $\Delta\Rey_{\tau}^{\mathrm{S}}$ and $\Delta\Rey_{\tau}^{\mathrm{H}}$ for (\subref{fig:std_turb_re5k}) case A, and (\subref{fig:std_turb_re8k}) case B.
    }
    \label{fig:std_turb}%
\end{figure}

We eventually show turbulent space-time diagrams of the excess friction Reynolds numbers relative to the laminar base flow solution,
\begin{subequations}
  \label{eq:excess_retau}%
  \begin{gather}
    \Delta\Rey_{\tau}^{\mathrm{S}}(z,t) = \Rey_{\tau}^{\mathrm{S}}(z,t) - \widetilde{\Rey}_{\tau}^{\mathrm{S}}(z), \qquad 
    \Delta\Rey_{\tau}^{\mathrm{H}}(y,t) = \Rey_{\tau}^{\mathrm{H}}(y,t) - \widetilde{\Rey}_{\tau}^{\mathrm{H}}(y),
    \tag{\theequation a,b}
  \end{gather}
\end{subequations}
for case A and B.
(Note that these quantities are time dependent and that the time averaging operator in (\ref{eq:turbulence_quantities}a) is omitted.)
This simplifies the detection of laminar patches, which in Fig.~\ref{fig:std_turb} appear as light regions, whereas the dark areas correspond to turbulence. 
Significant turbulent activity is recorded on the Shercliff wall for both flow configurations, 
whereas the turbulence on the Hartmann wall for the case A is largely confined to the corner region at $y\gtrsim 1-1/\sqrt{Ha}=0.77$.
It is emphasized that this localization of turbulence to one wall only is an MHD effect due to the damping by the Lorentz force. It should not be attributed to the domain length $L$, as reported for channel flow in \cite{jimenez_moin_1991}, nor to the symmetry relations \eqref{eq:symrel_vel} and \eqref{eq:symrel_cur}.
Evidence that this is the case is provided in Appendix \ref{sec:turb_stat}, where consistent localization is found for both the quarter and the full duct of length $L=8\pi$.
The $y$-location of maximum excess friction Reynolds number in Fig.~\ref{fig:std_turb_re5k} compares well with the location $y = 0.95$ (in outer units) of the peak of the turbulent fluctuation intensity in Fig.~\ref{fig:stat_uu}.
Relative to case A, the turbulent activity spreads over large portions of the Hartmann wall as the Reynolds number is increased to $R=400$ (case B), in line with the findings of Ref.~\cite{moresco2004experimental,krasnov2004numerical}.
The spatial structures become more fine-grained, and fluctuate more rapidly in time.

%=============================================================================
\section{Transition routes}
\label{sec:transition}%

\subsection{Transition process starting from localized initial conditions \label{sec:3a}}

Our aim is now to characterize the transition process, with an emphasis on the location in the duct cross-section where turbulent fluctuations first appear. 
To this end, we define the local perturbation enstrophy density $\mathcal{Z}$ by
\begin{equation}
    \mathcal{Z}=\frac{1}{2}||{\bm \omega}||^2,
    \label{eq:Z}%
\end{equation}
where ${\bm \omega}={\bm \nabla}\times(\vv - \widetilde{v}_x\ee_x)$ is the perturbation vorticity and $\|\cdot\|$ denotes the Euclidean norm.
Next, we define the $\mathcal{Z}$-weighted volumetric averages of the original coordinates $y$ and $z$ according to
\begin{subequations}
  \label{eq:enstrophy_centroid}%
  \begin{gather}
    \hat{y}=\frac{\int_{\Omega}y\mathcal{Z}\,\dd\upsilon}{\int_{\Omega}\mathcal{Z}\,\dd\upsilon}, \qquad
    \hat{z}=\frac{\int_{\Omega}z\mathcal{Z}\,\dd\upsilon}{\int_{\Omega}\mathcal{Z}\,\dd\upsilon}.
    \tag{\theequation a,b}
  \end{gather}
\end{subequations}
The quantities $\hat{y}$ and $\hat{z}$ are interpreted as the instantaneous coordinates, in a cross-section of the duct, of the barycenter of the perturbation enstrophy field. 
During transition to turbulence, this point is expected to shift towards the boundary layer where enstrophy growth is most intense.
Note that for a space-filling turbulent flow field, $\hat{y}=\hat{z}=0.5$ in the present quarter duct.
The expressions for $\hat{y}$ and $\hat{z}$ can be readily compared with those for the center of vorticity (\cf~\S7.3 in \cite{batchelor1967introduction}, with enstrophy replaced by vorticity) and are here interpreted in an analogous fashion.

\begin{figure}
    \centering
    \begin{subfigure}[b]{0.495\textwidth}
        \setlength{\unitlength}{1.0cm}
        \centering
        \begin{picture}(8.4,7.5)
            \put(-0.3,0.0){\includegraphics[scale=0.5]{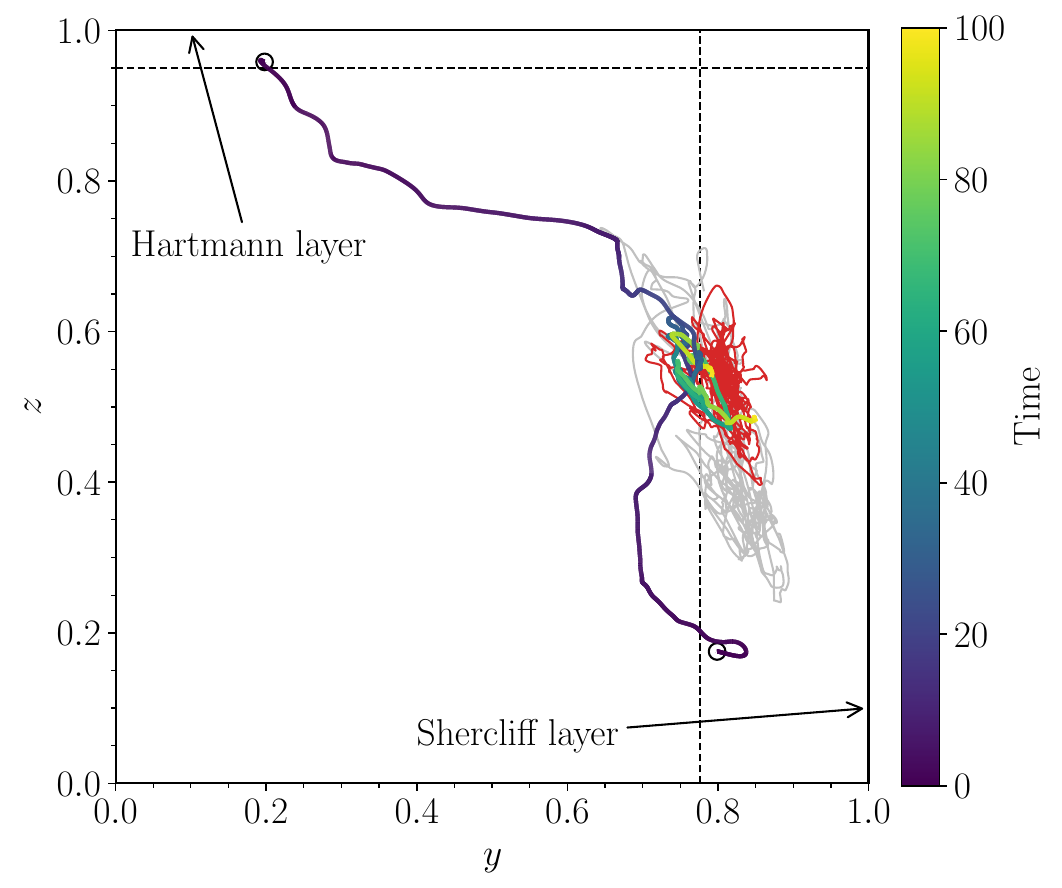}}%
            \put(-0.2,0.0){(a)}
        \end{picture}
        \phantomsubcaption
        \label{fig:enstrophy_centroid_re5k}%
    \end{subfigure}          
    \begin{subfigure}[b]{0.495\textwidth}
        \setlength{\unitlength}{1.0cm}
        \centering
        \begin{picture}(8.4,7.5)
            \put(-0.3,0.0){\includegraphics[scale=0.5]{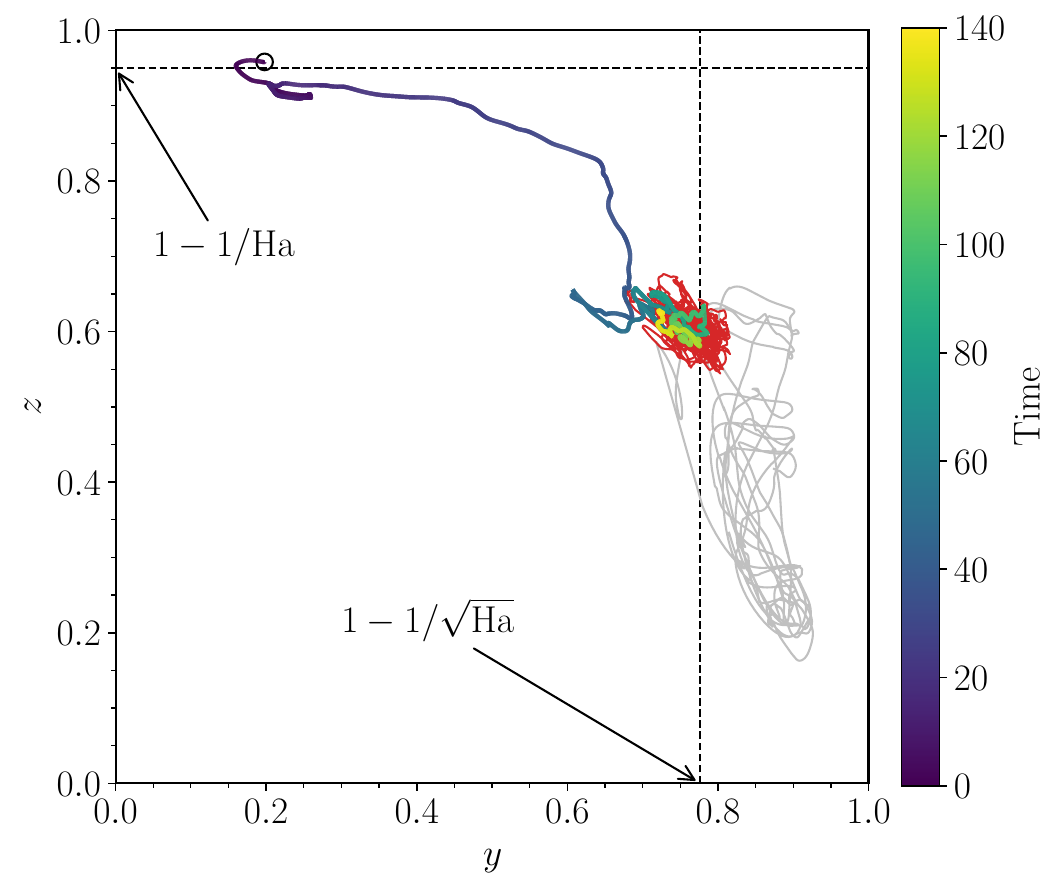}}%
            \put(-0.2,0.0){(b)}
        \end{picture}
        \phantomsubcaption
        \label{fig:enstrophy_centroid_re8k}%
    \end{subfigure}          
    \caption{
    Evolution of the enstrophy centroid $(\hat{y},\hat{z})$ for perturbations initially placed in the Shercliff or the Hartmann layer.
    The beginning of each trajectory is marked with a small black circle, and the color of the lines change as the time progresses.
    For comparison, corresponding data for turbulent flow (gray) and the edge state (red, see \S\ref{sec:edge_state}).
    As reference, the approximate thicknesses of the Shercliff and the Hartmann layers are indicated with dashed lines. 
    (\subref{fig:enstrophy_centroid_re5k}) case A, (\subref{fig:enstrophy_centroid_re8k}) case B.
    }
    \label{fig:enstrophy_centroid}%
\end{figure} 

To initiate transition independently in the Shercliff and the Hartmann layer, the two boundary layers are individually perturbed by the synthetic disturbance described in Appendix \ref{sec:synthpert}.
The thickness of the Shercliff and the Hartmann layer scales as $\delta^{\mathrm{S}}\sim \Ha^{-1/2}$ and $\delta^{\mathrm{H}}\sim\Ha^{-1}$ \cite{zikanov2014laminar}, respectively.
We choose here unit proportionality constants and center the perturbations at distances $\Ha^{-1/2}$ and $\Ha^{-1}$ from either wall.
The time evolution of $\hat{y}$ and $\hat{z}$ is plotted in Fig.~\ref{fig:enstrophy_centroid_re5k} for case A. 
Regardless of where the perturbation is positioned initially, the vortical motion migrates towards the center of the Shercliff wall in the quarter duct where it fluctuates at a distance of $\approx\Ha^{-1/2}$ from the wall.
One can compare the terminal position of these trajectories with the corresponding values of $\hat{y}$ and $\hat{z}$ for fully developed turbulence: they roughly occupy the same spatial regions.

\begin{figure}
    \centering
    \begin{subfigure}[b]{0.52\textwidth}
        \setlength{\unitlength}{1.0cm}
        \centering
        \begin{picture}(8.6,7.0)
            \put(-0.1,0.0){\includegraphics[scale=0.5]{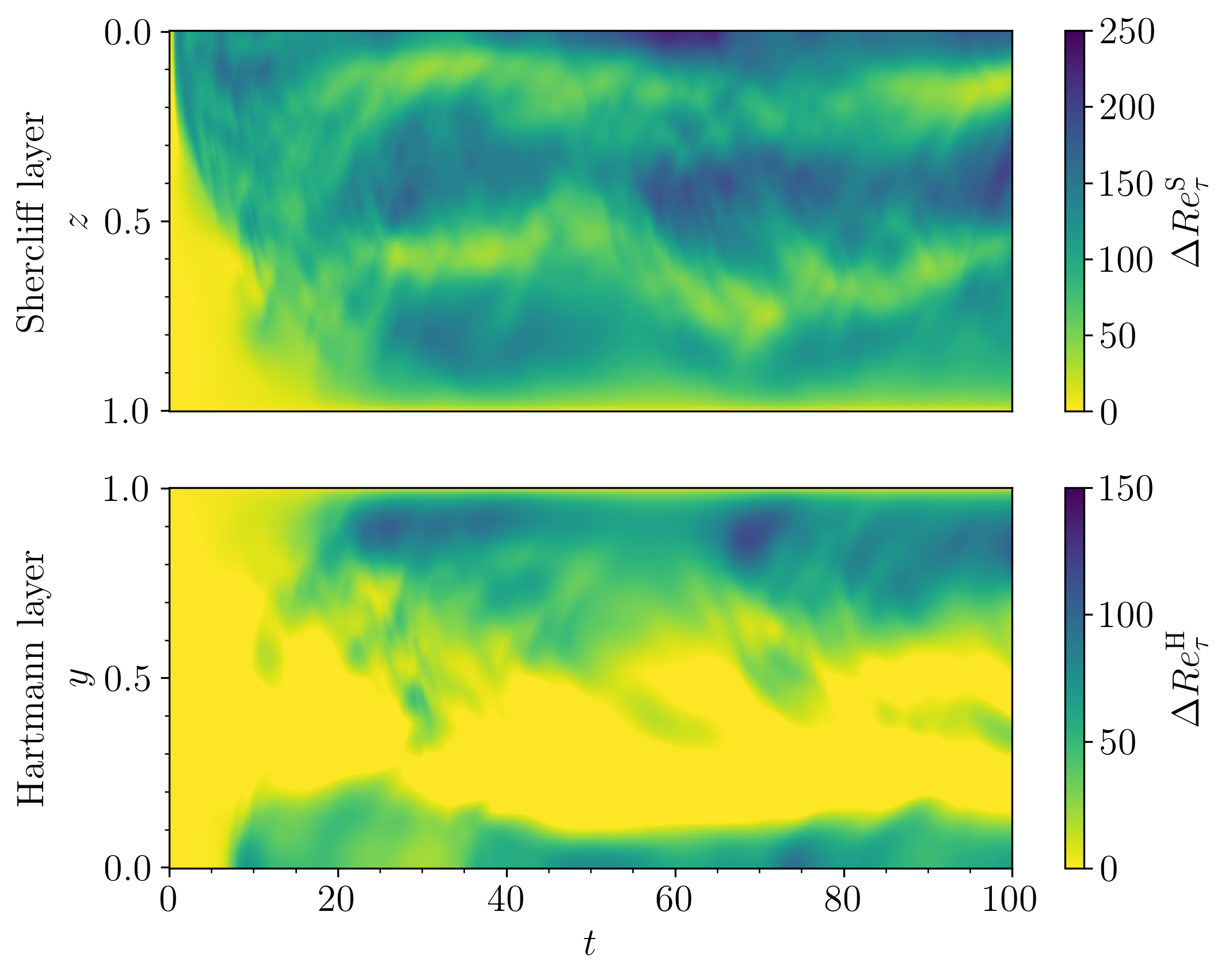}}%
            \put(-0.1,0.0){(a)}
        \end{picture}      
        \phantomsubcaption
        \label{fig:std_trans_shercliff_re5k}%
    \end{subfigure}          
    \begin{subfigure}[b]{0.47\textwidth}
        \setlength{\unitlength}{1.0cm}
        \centering
        \begin{picture}(7.7,7.0)
            \put(-0.2,0.0){\includegraphics[scale=0.5]{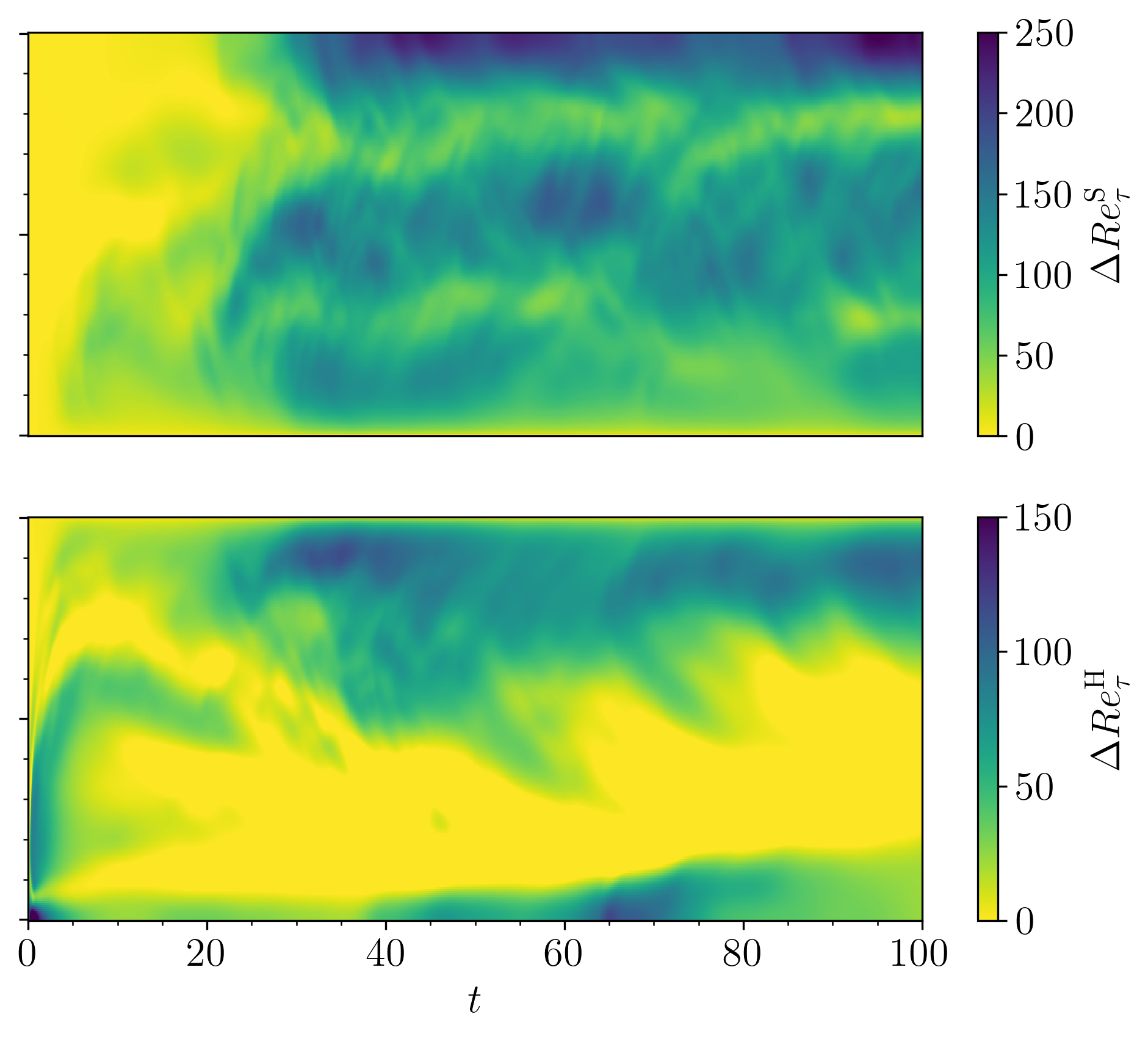}}%
            \put(-0.2,0.0){(b)}
        \end{picture}
        \phantomsubcaption
        \label{fig:std_trans_hartmann_re5k}%
    \end{subfigure}
    \caption{Space-time diagrams of excess friction Reynolds number $\Delta\Rey_{\tau}^{\mathrm{S}}$ and $\Delta\Rey_{\tau}^{\mathrm{H}}$
    for case A.
    Transition triggered by a perturbation localized in (\subref{fig:std_trans_shercliff_re5k}) the Shercliff layer, or in (\subref{fig:std_trans_hartmann_re5k}) the Hartmann layer.
    }
    \label{fig:std_trans_re5k}%
\end{figure}

An alternative view of studying the motion of the perturbation is through space-time diagrams, as shown in Fig.~\ref{fig:std_trans_re5k}.
In Fig.~\ref{fig:std_trans_shercliff_re5k}, the spreading of the perturbation across the Shercliff wall is clearly visible, and it is completed in approximately 20 advective time units.
This explains why $\hat{z}\approx 0.5$ for long times in Fig.~\ref{fig:enstrophy_centroid_re5k}.
Similarly, the immediate migration of the perturbation from the Hartmann to the Shercliff layer is visible in Fig.~\ref{fig:std_trans_hartmann_re5k}.
This process is seen to be completed in about 30 advective time units.

As $\Rey$ is increased from 5000 to 8000 (\cf~case A and case B), the situation is largely unchanged.
This is shown in Fig.~\ref{fig:enstrophy_centroid_re8k} for case B, where only the Hartmann layer is perturbed.
A space-time diagram similar to the one presented in Fig.~\ref{fig:std_trans_hartmann_re5k} (not shown) gives that the transition to turbulence has spread to the Shercliff layer about 40 time units after it was incited in the Hartmann layer.
Although Fig.~\ref{fig:std_turb_re8k} shows that the Hartmann wall is capable of supporting turbulence, a flow featuring a turbulent Hartmann wall together with a laminar Shercliff wall appears to be ruled out.

\subsection{Edge states}
\label{sec:edge_state}%

\begin{figure}
    \centering
    \begin{subfigure}[b]{0.49\textwidth}
        \setlength{\unitlength}{1.0cm}
        \centering
        \begin{picture}(8.25,4.0)
            \put(-0.06,0.0){\includegraphics[scale=0.5]{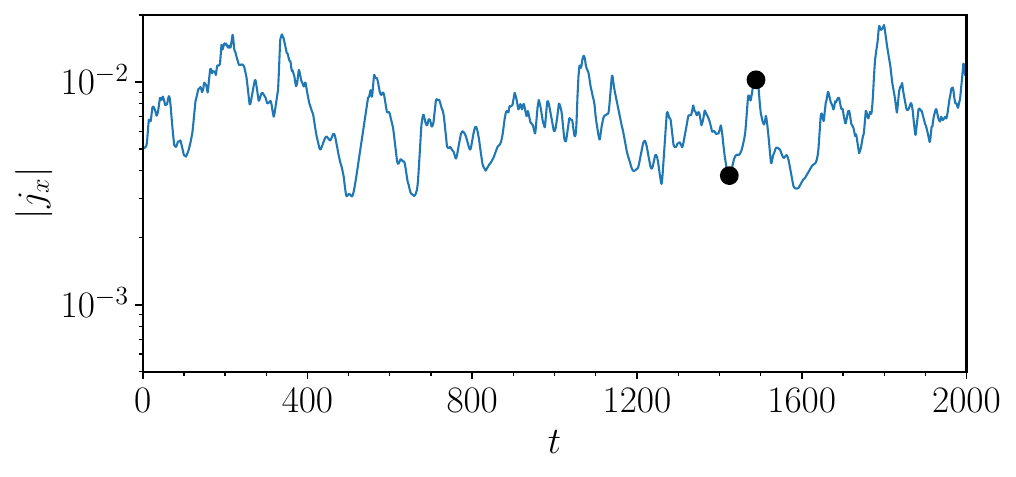}}%
            \put(-0.1,0.0){(a)}
        \end{picture}
        \phantomsubcaption
        \label{fig:edge_traj_re5k}%
    \end{subfigure}          
    \begin{subfigure}[b]{0.49\textwidth}
        \setlength{\unitlength}{1.0cm}
        \centering
        \begin{picture}(8.25,4.0)
            \put(-0.06,0.0){\includegraphics[scale=0.5]{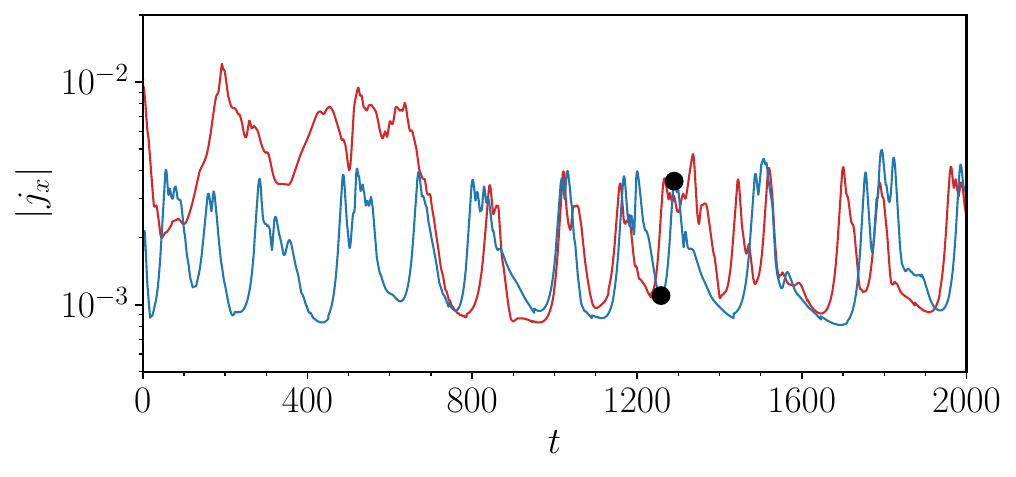}}%
            \put(-0.1,0.0){(b)}
        \end{picture}
        \phantomsubcaption
        \label{fig:edge_traj_re8k}%
    \end{subfigure}  
    \caption{ 
    Time series of $|j_x|$ corresponding to edge trajectories for (\subref{fig:edge_traj_re5k}) case A and (\subref{fig:edge_traj_re8k}) case B.
    In (\subref{fig:edge_traj_re8k}), edge trajectories resulting from an initial perturbation either in the Hartmann layer (red) or in the Shercliff layer (blue) are shown.
    Consequently, while the transient phase is retained in (\subref{fig:edge_traj_re8k}), it has been truncated in (\subref{fig:edge_traj_re5k}) with an appropriate shift of the time variable.
    The black circles correspond to the snapshots visualized in Fig.~\ref{fig:edgestate_re5k} and Fig.~\ref{fig:edgestate_re8k}, respectively. 
    }
    \label{fig:edge_traj}%
\end{figure} 

Since the turbulent regimes of interest are nonlinear equilibrium regimes, these observations call for an explanation of a nonlinear nature as well.  The identification of relevant nonlinear equilibrium solutions to the governing equations, hopefully simpler to interpret than the very complex turbulent regimes, is a possible way to explain or at least justify the observed turbulent dynamics. 
Such nonlinear solutions may take the shape of fixed points, traveling waves, periodic orbits or relative periodic orbits possibly interconnected by heteroclinic and/or homoclinic orbits.
Since the significance of such exact coherent structures for turbulent flow may sometimes be dubious, we chose to focus on edge states, which are well defined and relevant to the subcritical situation where the laminar and turbulent regime coexist.
In spite of being an edge state, the determined solutions may still be classified as one of the above mentioned types.

Edge states correspond to unstable flow regimes distinct from the laminar and the turbulent state. They have the defining property that they lie on the basin boundary of the laminar regime, the so-called \emph{edge of chaos} \citep{skufca2006edge,schneider2007turbulence}. They are interpreted as key mediators of the transitional dynamics and as the first nonlinear coherent structures visited along a transition route \cite{khapko2016edge}. The temporal dynamics of edge states can be simple or complex depending on conditions not fully understood yet. However, in this article we are less interested in their temporal dynamics than in their spatial localization properties, in particular whether a given edge state solution can be associated with one given duct wall. Competition between several edge state solutions of a physical system has been reported several times in the literature, \eg~in \cite{khapko2014complexity}. No general rule gives any information about the unicity of edge state solutions. The conclusions given in this section are therefore based on computational evidence.

\begin{figure}
    \centering
    \begin{subfigure}[b]{0.48\textwidth}
        \setlength{\unitlength}{1.0cm}
        \centering
        \begin{picture}(7.0,9.0)
            \put(0.0,0.0){\includegraphics[width=0.8\textwidth]{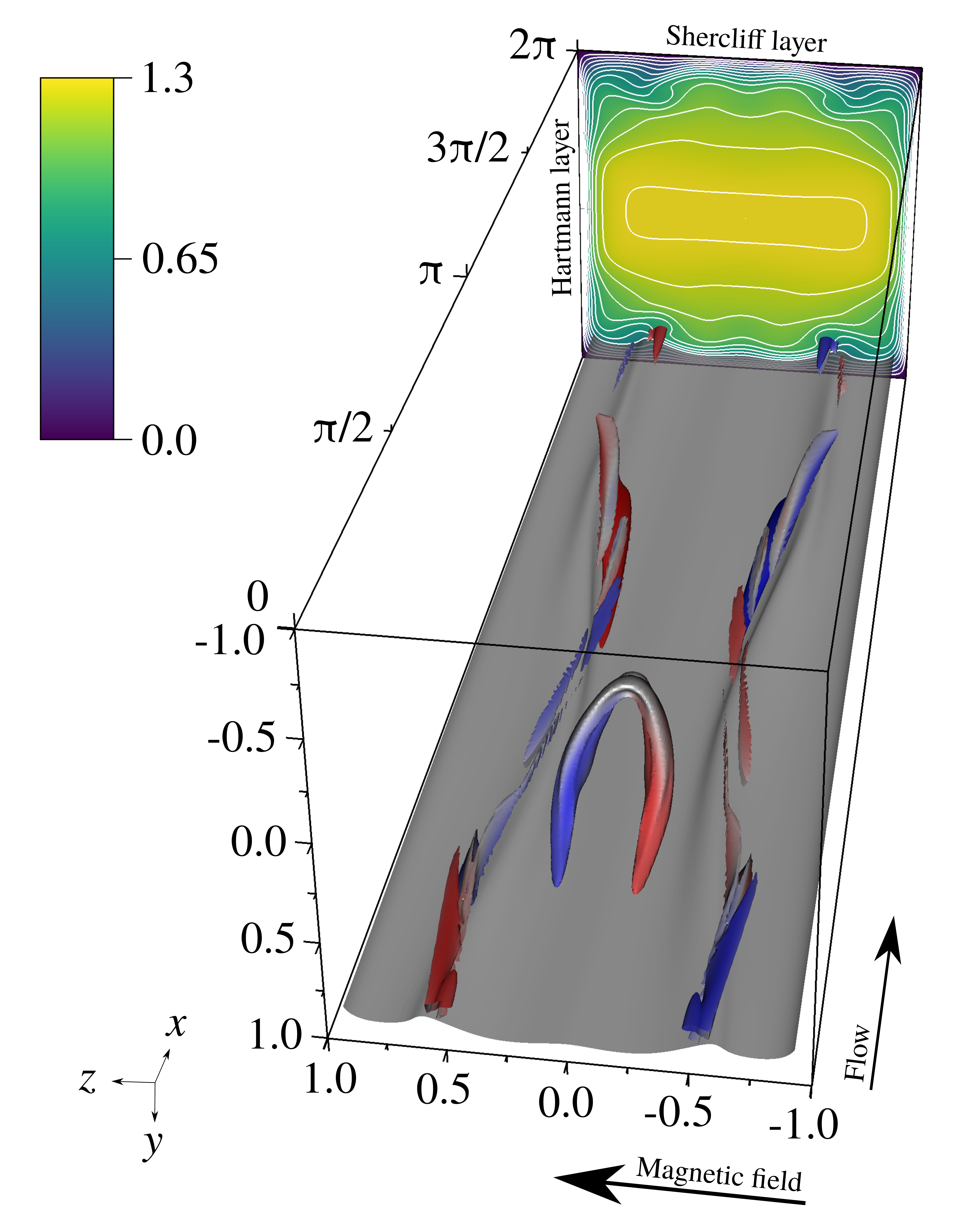}}%
            \put(-0.1,0.1){(a)}
        \end{picture}
        \phantomsubcaption
        \label{fig:edgestate_re5k_calm}%
    \end{subfigure}
    \begin{subfigure}[b]{0.48\textwidth}
        \setlength{\unitlength}{1.0cm}
        \centering
        \begin{picture}(7.0,9.0)
            \put(0.0,0.0){\includegraphics[width=0.8\textwidth]{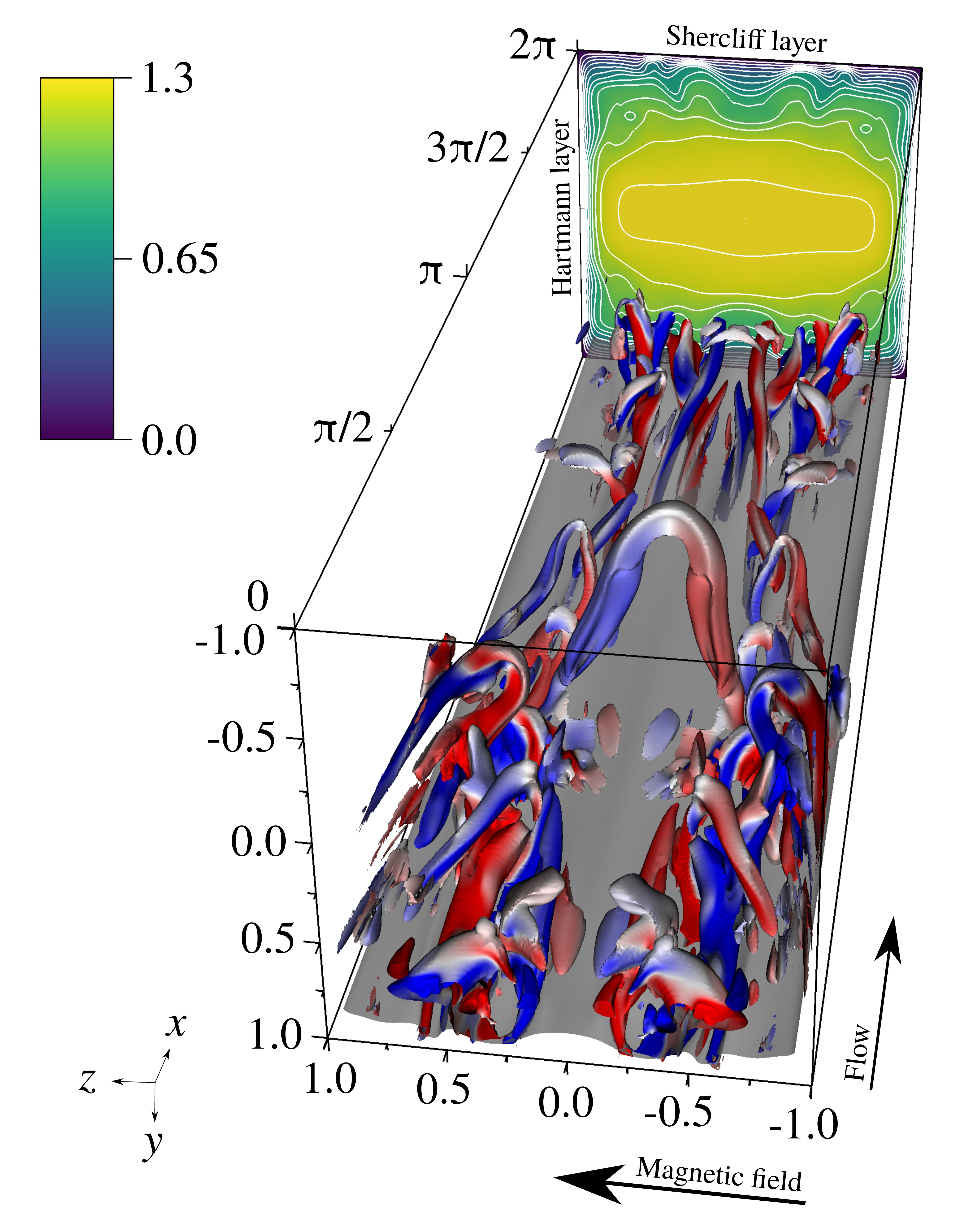}}%
            \put(-0.1,0.1){(b)}
        \end{picture}
        \phantomsubcaption
        \label{fig:edgestate_re5k_intense}%
    \end{subfigure}   
    \caption{Instantaneous visualizations of the edge state for case A during (\subref{fig:edgestate_re5k_calm}) a calm, and (\subref{fig:edgestate_re5k_intense}) an intense phase.
    Isocontours of streamwise velocity $v_x = 0.6$ (gray), together with a transversal plane of $v_x$ (color, the white lines are separated by 0.1 units).
    Vortical structures are shown using an isocontour of $\lambda_2=-0.03$ \cite{jeong1995identification} colored by streamwise vorticity $\omega_x$ (red - positive, blue - negative).
    The time of the snapshots correspond to $t=1\,424.2$ and $t=1\,489.0$ as indicated in Fig.~\ref{fig:edge_traj_re5k}.    
    For visualization purposes, the quarter duct domain is extended to the full duct.
    }
    \label{fig:edgestate_re5k}%
\end{figure}

\begin{figure}
    \centering
    \begin{subfigure}[b]{0.48\textwidth}
        \setlength{\unitlength}{1.0cm}
        \centering
        \begin{picture}(7.0,9.0)
            \put(0.0,0.0){\includegraphics[width=0.8\textwidth]{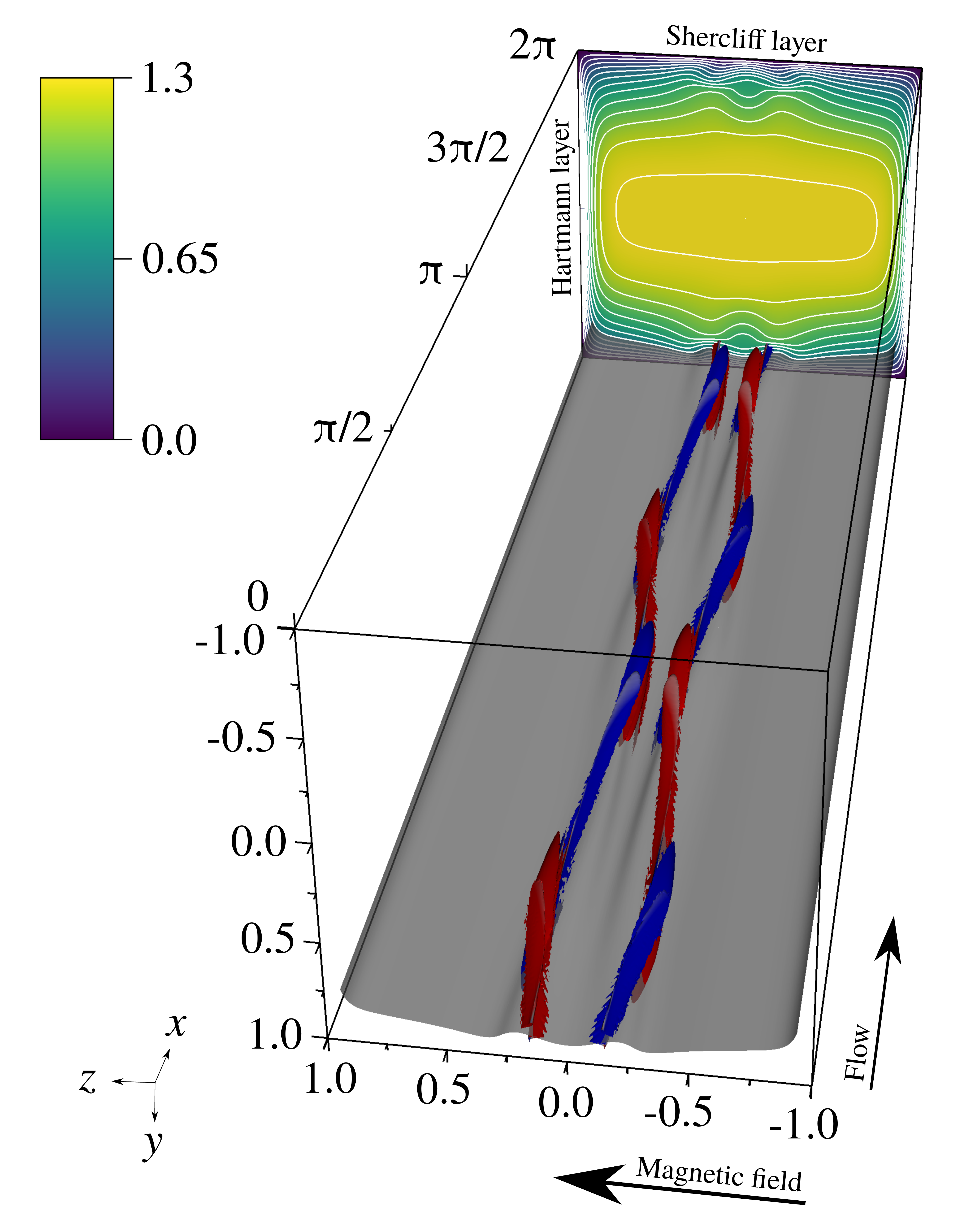}}%
            \put(-0.1,0.1){(a)}
        \end{picture}
        \phantomsubcaption
        \label{fig:edgestate_re8k_calm}%
    \end{subfigure}
    \begin{subfigure}[b]{0.48\textwidth}
        \setlength{\unitlength}{1.0cm}
        \centering
        \begin{picture}(7.0,9.0)
            \put(0.0,0.0){\includegraphics[width=0.8\textwidth]{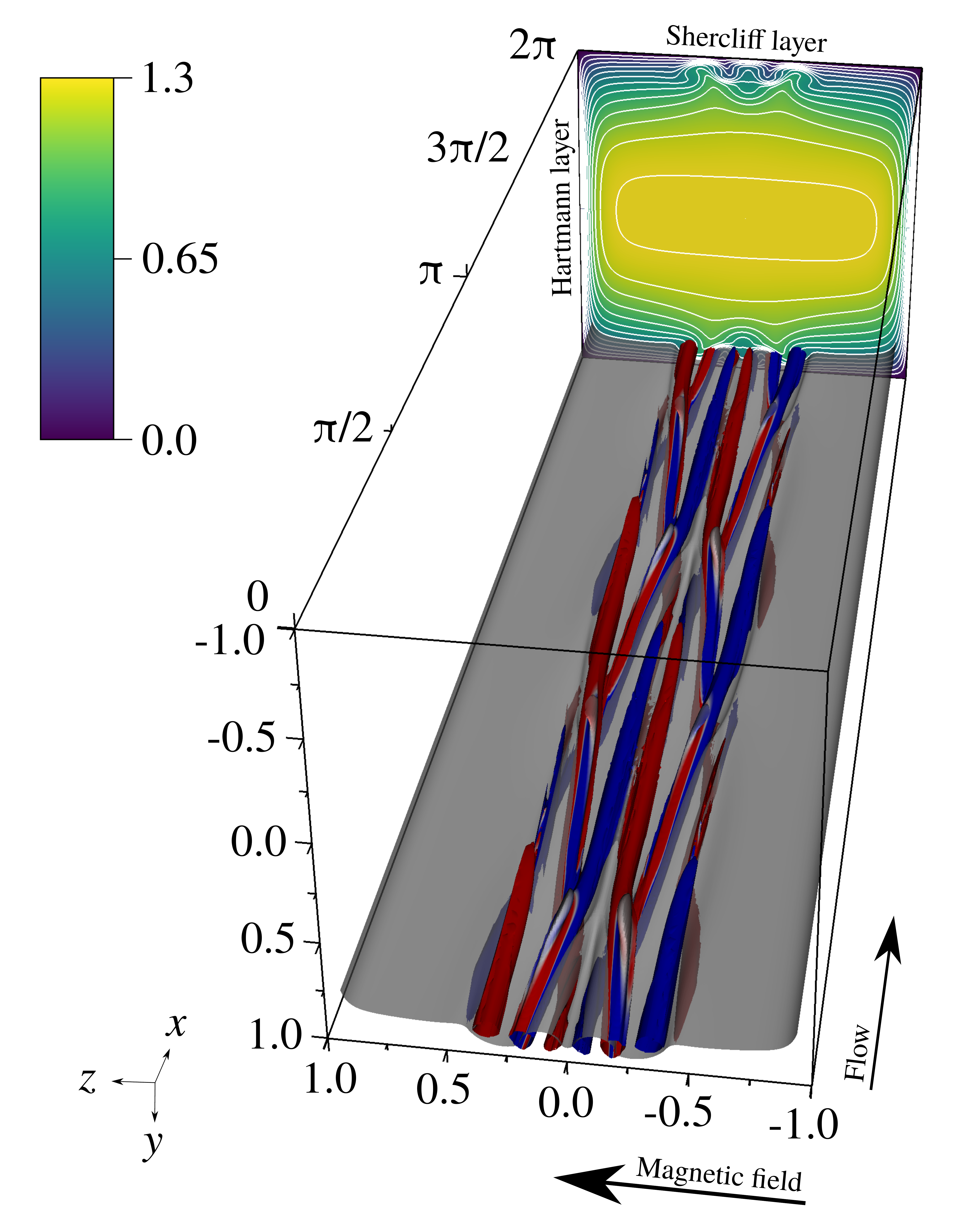}}%
            \put(-0.1,0.1){(b)}
        \end{picture}
        \phantomsubcaption
        \label{fig:edgestate_re8k_burst}%
    \end{subfigure}   
    \caption{Instantaneous visualizations of the edge state for case B during (\subref{fig:edgestate_re8k_calm}) a calm, and (\subref{fig:edgestate_re8k_burst}) a bursting phase.
    Isocontours of streamwise velocity $v_x = 0.6$ (gray), together with a transversal plane of $v_x$ (color, the white lines are separated by 0.1 units).
    Vortical structures are shown using an isocontour of $\lambda_2=-0.01$ \cite{jeong1995identification} colored by streamwise vorticity $\omega_x$ (red - positive, blue - negative).
    The time of the snapshots correspond to $t=1\,258.4$ and $t=1\,290.4$ as indicated
    in Fig.~\ref{fig:edge_traj_re8k}. 
    For visualization purposes, the quarter duct domain is extended to the full duct
    and two streamwise periods are shown.
    }
    \label{fig:edgestate_re8k}%
\end{figure}

A popular method to determine an edge state is bisection \cite{itano2001dynamics,toh2003periodic,skufca2006edge}. 
The algorithm begins by the choice of two initial conditions $(\vv^{\mathrm{lam}},\vv^{\mathrm{turb}})$ belonging respectively to the laminar and the turbulent attraction basin. The segment joining these two initial conditions must cross the edge manifold at least once. 
An initial condition $\vv^{\mathrm{bis}}$ belonging to the edge is identified by iteratively updating the weight coefficient $\varsigma$ in
\begin{equation}
  \vv^{\mathrm{bis}}(\xv) = \varsigma \vv^{\mathrm{lam}}(\xv) + (1-\varsigma)\vv^{\mathrm{turb}}(\xv),
  \label{eq:bisection}%
\end{equation}
where $0<\varsigma<1$.
The edge state is eventually identified as the long-time attractor for the trajectory initiated by the initial condition on the edge manifold. In practice however, neighboring trajectories separate exponentially fast in the direction transverse to the edge \cite{schneider2007turbulence,brynjell-rahkola_etal_2024} due to the positive Lyapunov exponent. The bisection needs thus to be repeatedly restarted until convergence is achieved. As a diagnostic for turbulence transition or laminarization, we monitor here at all times the norm of the streamwise current density $|j_x|$ defined in \eqref{eq:jx_norm}, which vanishes for the base state.  For each pair of initial conditions, bisection is performed until the distance between the two states, as measured by the $L^2$-norm, falls below $10^{-5}$. A new bisection is initiated when the separation between the trajectories has reached $10^{-3}$ (as measured in the $L^2$-norm).

Starting with case A, a time series of an edge trajectory is shown in Fig.~\ref{fig:edge_traj_re5k}. 
As in HD duct and pipes \cite{schneider2007turbulence,duguet2008transition,biau2009optimal}, it displays chaotic dynamics. However, whereas \cite{biau2009optimal} reported that different duct walls become active at different times (suggesting that the edge state cycles around the duct circumference in time), the equivalence between all walls is lost in the presence of a magnetic field.
Instead, the edge state stays localized on the Shercliff wall.
Two instantaneous snapshots of the edge state at arbitrary times (see Fig.~\ref{fig:edge_traj_re5k}) corresponding to calm and intense episodes are shown in Fig.~\ref{fig:edgestate_re5k}.
Whereas the calm phase exhibits a remarkable degree of coherence and order, these structures are seen to break down into mild turbulence during the active stages.
Similar behavior was recently reported in \cite{brynjell-rahkola_etal_2024} in the MHD channel.
Notable in both visualizations is the presence of a large hairpin vortex next to one of the symmetry lines. Whether this feature is due to the imposed boundary conditions is presently not known. Hairpin vortices are however known to develop in the early stages of boundary layer transition and promote turbulent spot nucleation \cite{schlatter_2012turbulent}.
If the flow would be perturbed along the unstable direction of the edge state transverse to the edge manifold, it is likely that turbulence would ensue following a similar breakdown of this hairpin into smaller vortices.

The particular trajectory shown in \ref{fig:edge_traj_re5k} corresponds to a bisection initiated between a pair of snapshots $(\vv^{\mathrm{lam}},\vv^{\mathrm{turb}})$ taken to be the laminar solution shown in Fig.~\ref{fig:baseflow} and a low-intensity turbulent state.
To check the robustness of the results, the bisection was repeated by choosing $\vv^{\mathrm{turb}}$ to be the initial perturbation described in Appendix \ref{sec:synthpert} centered on the Shercliff wall, or another turbulent state but using slightly different solver tolerances (see \cite{nek5000} for details).
All these cases converge to similar results (as far as chaotic states can be quantitatively compared based on limited time series).

\begin{figure}
    \centering
    \begin{subfigure}[b]{0.52\textwidth}
        \setlength{\unitlength}{1.0cm}
        \centering
        \begin{picture}(8.6,7.0)
            \put(-0.2,0.0){\includegraphics[scale=0.5]{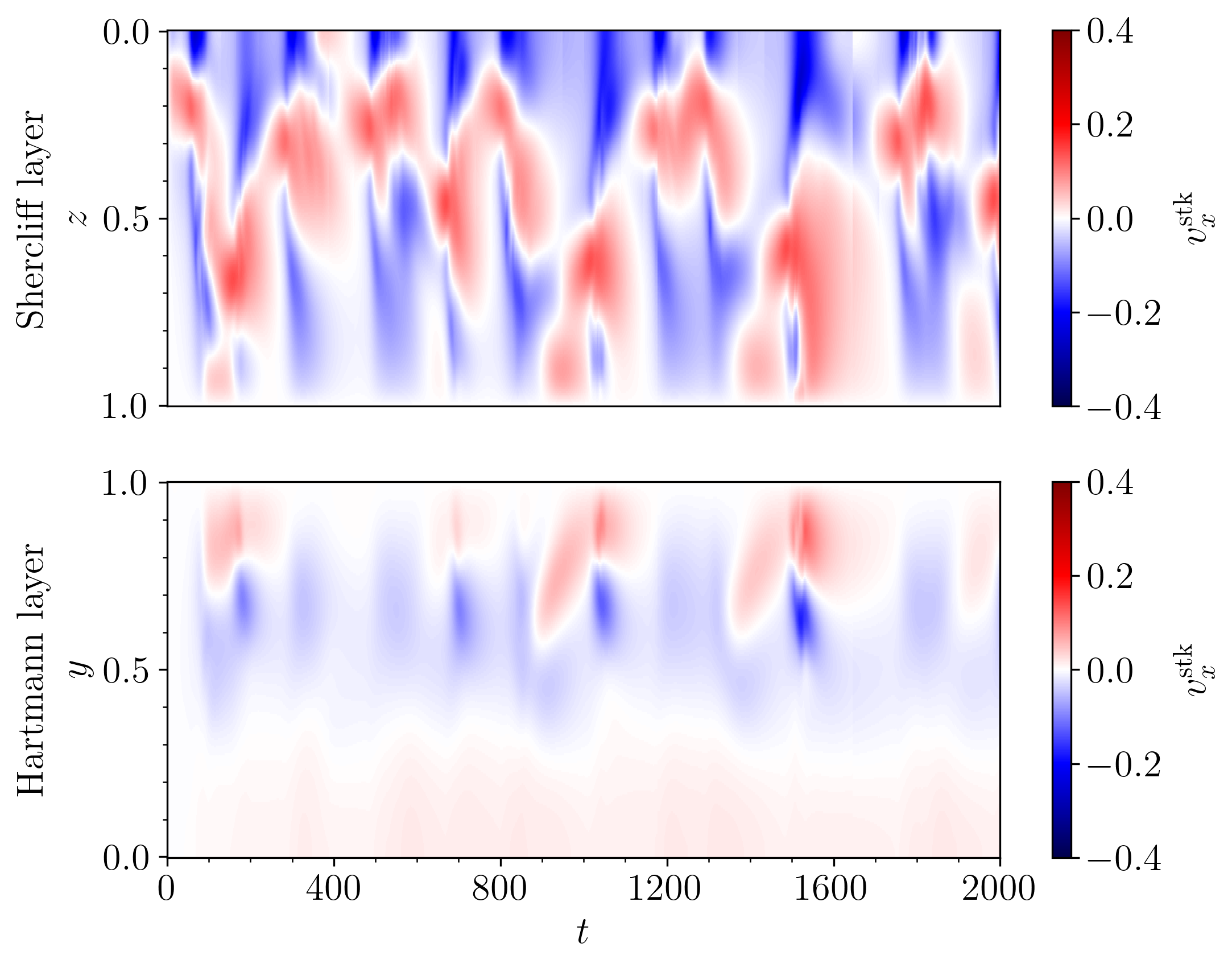}}%
            \put(-0.2,0.0){(a)}
        \end{picture}       
        \phantomsubcaption
        \label{fig:std_edge_shercliff}%
    \end{subfigure}          
    \begin{subfigure}[b]{0.47\textwidth}
        \setlength{\unitlength}{1.0cm}
        \centering
        \begin{picture}(7.7,7.0)
            \put(-0.1,0.0){\includegraphics[scale=0.5]{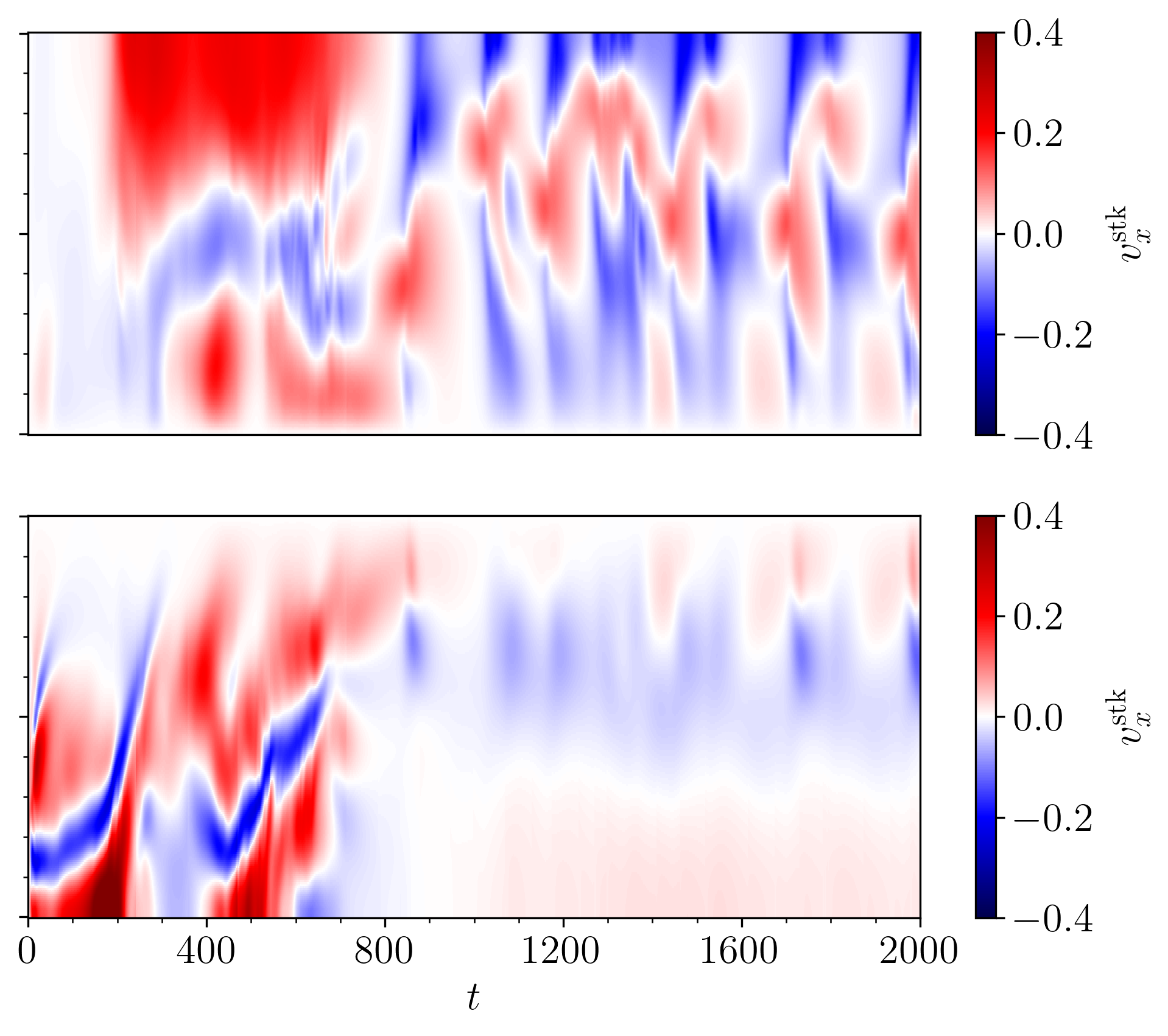}}%
            \put(-0.1,0.0){(b)}
        \end{picture}
        \phantomsubcaption
        \label{fig:std_edge_hartmann}%
    \end{subfigure}
    \caption{Space-time diagrams of the edge state in case B visualized using the streamwise velocity streak $v_x^{\mathrm{stk}}$ along the dashed lines indicated in Fig.~\ref{fig:enstrophy_centroid_re8k}, \ie~$y=1-1/\sqrt{\Ha}$ (upper panels) and $z=1-1/\Ha$ (lower panels).
    Bisection is initiated using a perturbation in (\subref{fig:std_edge_shercliff}) the Shercliff layer and (\subref{fig:std_edge_hartmann}) the Hartmann layer.
    }
    \label{fig:std_edge}%
\end{figure}

Case B is considered next.
Since sustained turbulent motion was shown on the Hartmann wall in Fig.~\ref{fig:std_turb_re8k} at this parameter value,
two separate bisections are initiated between the laminar solution $\vv^{\mathrm{lam}}$ and the disturbance of Appendix \ref{sec:synthpert}, localized now either at the Shercliff or at the Hartmann wall as $\vv^{\mathrm{turb}}$.
In Fig.~\ref{fig:edge_traj_re8k}, the signal of the edge trajectory initiated from a perturbation in the Shercliff layer is qualitatively different from that observed for case A.
Instead of being fully chaotic, the edge state now exhibits a recurrent cycle of quiescent phases interrupted by rapid bursts reminiscent to that previously reported for HD Poiseuille flow \cite{itano2001dynamics,zammert2014periodically,neelavara2017state}, asymptotic suction boundary layer \cite{kreilos2013edge,khapko2013localized} and the MHD channel flow for low to intermediate $\Ha$-numbers \cite{brynjell-rahkola_etal_2024}.
Instantaneous snapshots of the edge state during the quiescent and the bursting stage indicated in Fig.~\ref{fig:edge_traj_re8k} are shown in Fig.~\ref{fig:edgestate_re8k}.
The space-time diagram in Fig.~\ref{fig:std_edge_shercliff} shows that the low- and high-speed streaks defined by $v_x^{\mathrm{stk}}=v_x - \widetilde{v}_x$ shift erratically along the Shercliff wall in time, similarly to the case of intermediate $\Ha$ in the MHD channel flow (\cf~Fig.~\ref{fig:edge_traj_re8k} and \ref{fig:std_edge_shercliff} with figure 3c and 5b of \cite{brynjell-rahkola_etal_2024}).
Meanwhile, the Hartmann layer remains largely unperturbed as for lower $R$.

For the bisection initiated in the Hartmann layer, Fig.~\ref{fig:std_edge_hartmann} shows that the edge dynamics remains isolated in this layer until time $t=200$ when it migrates into the Shercliff layer. Initially the level of $|j_x|$ is high and the time series do not show any sign of recurrence (see Fig.~\ref{fig:edge_traj_re8k}). From $t=700$ onward, the migration to the Shercliff wall is complete and the dynamics converge to the recurrent cycle. Beyond $t=700$, the flow next to the Hartmann wall returns to its nearly unperturbed state.
Hence, this rules out edge states that would be either localized to the Hartmann layer or not localized at all.

\subsection{State space projections}
\label{sec:3c}%

Fig.~\ref{fig:enstrophy_centroid} shows the physical location of the main vortical motion. It suggests that both transition trajectories are attracted to the same physical region as occupied by the edge state. It is of complementary interest to visualize the trajectories using state portraits based on other physical quantities, notably energetic ones and wall shear stresses, since transition is associated with an increase in both of these. 
A popular choice of state variables 
stems from the energy balance
\begin{equation}
    \frac{\partial E}{\partial t} = I - D_{\nu} - D_{\mu},
    \label{eq:energy_balance}%
\end{equation}
where the different terms
\begin{subequations}
  \label{eq:budget_terms}%
  \begin{gather}
    E = 2\int_{\Omega} \vv\cdot\vv\,\dd \upsilon, \quad
    I = 4\int_{\Omega} \chi(\vv\cdot\ee_x)\,\dd \upsilon, \quad 
    D_{\nu} = \frac{4}{\Rey}\int_{\Omega} \nabla\vv:\nabla\vv\,\dd \upsilon, \quad
    D_{\mu} = \frac{4\Ha^2}{\Rey}\int_{\Omega} \jv\cdot\jv\,\dd \upsilon
    \tag{\theequation a,b,c,d}
  \end{gather}
\end{subequations}
respectively correspond to the kinetic energy of the flow, the energy input by the streamwise driving force, the viscous dissipation and the Joule dissipation linked to MHD effects.
(The terms are scaled to match the values in a full duct.)
In the case of a steady state such as the laminar solution, the left-hand side of \eqref{eq:energy_balance} vanishes such that $I=D_{\nu}+D_{\mu}$.
The laminar values of the dissipation terms (per unit length in the streamwise direction) in turn read 
$\widetilde{D}_{\nu}/L = 61.3354/\Rey$ and 
$\widetilde{D}_{\mu}/L = 43.0439/\Rey$.

\begin{figure}
    \centering
    \begin{subfigure}[b]{0.48\textwidth}
        \setlength{\unitlength}{1.0cm}
        \centering
        \begin{picture}(8.0,5.6)
            \put(0.38,0.0){\includegraphics[scale=0.5]{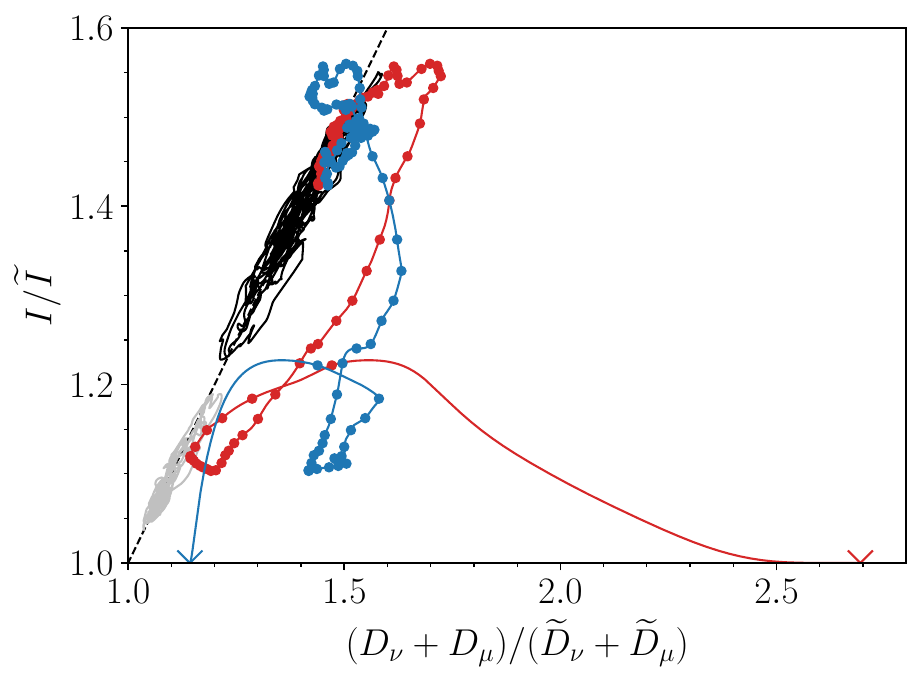}}%
            \put(0.0,0.1){(a)}
        \end{picture}
        \phantomsubcaption
        \label{fig:state_portrait_re5k_DI}%
    \end{subfigure}          
    \begin{subfigure}[b]{0.48\textwidth}
        \setlength{\unitlength}{1.0cm}
        \centering
        \begin{picture}(8.0,5.6)
            \put(0.29,0.0){\includegraphics[scale=0.5]{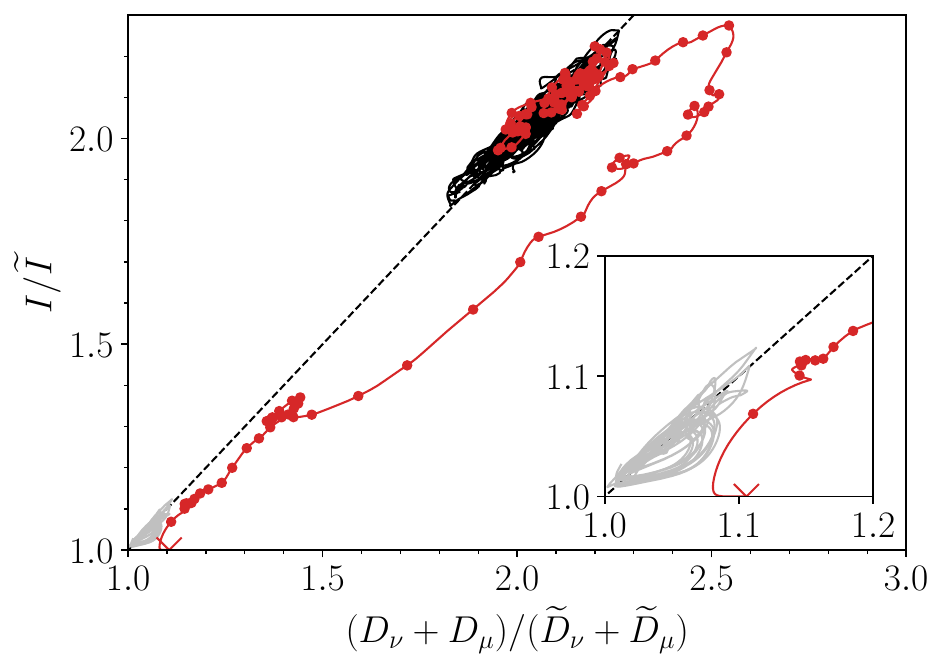}}%
            \put(0.0,0.1){(b)}
        \end{picture}
        \phantomsubcaption
        \label{fig:state_portrait_re8k_DI}%
    \end{subfigure}\\
    \vspace{1mm}
    \begin{subfigure}[b]{0.48\textwidth}
        \setlength{\unitlength}{1.0cm}
        \centering
        \begin{picture}(8.0,5.6)
            \put(0.2,0.0){\includegraphics[scale=0.5]{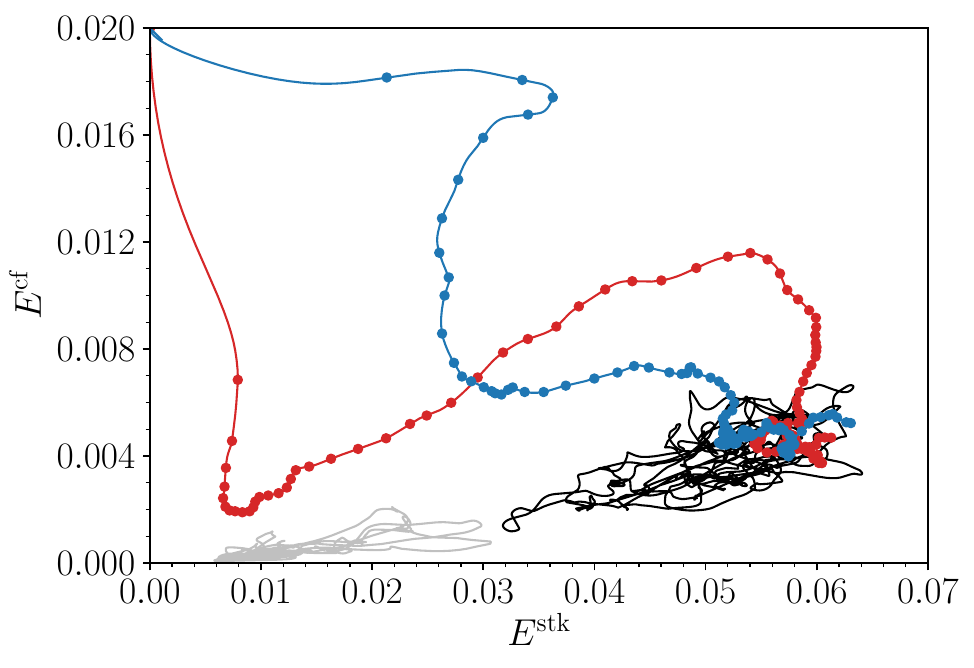}}%
            \put(0.0,0.1){(c)}
        \end{picture}
        \phantomsubcaption
        \label{fig:state_portrait_re5k_E}%
    \end{subfigure}
    \begin{subfigure}[b]{0.48\textwidth}
        \setlength{\unitlength}{1.0cm}
        \centering
        \begin{picture}(8.0,5.6)
            \put(0.2,0.0){\includegraphics[scale=0.5]{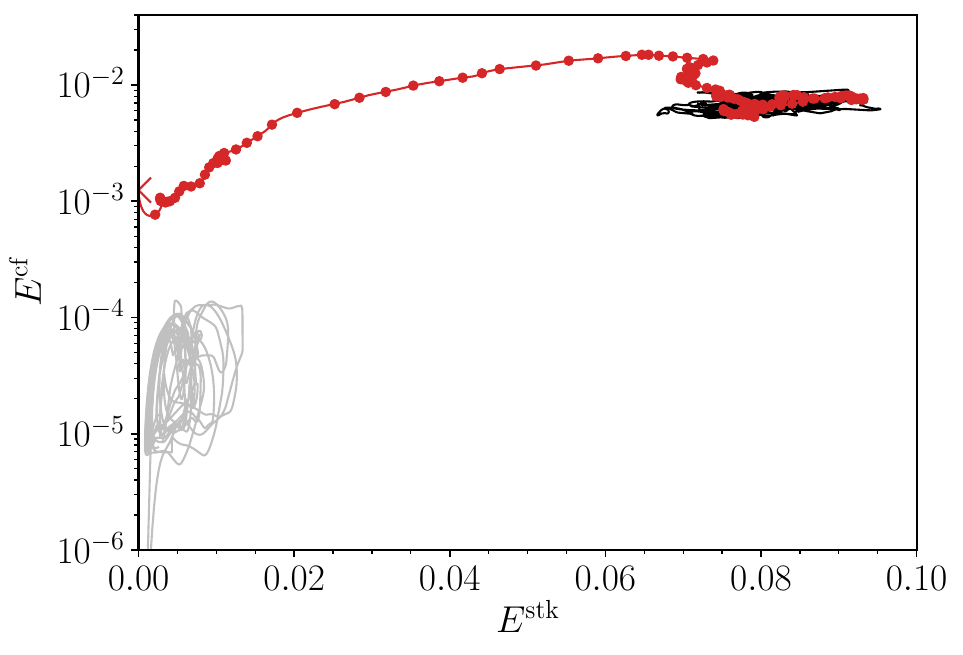}}%
            \put(0.0,0.1){(d)}
        \end{picture}
        \phantomsubcaption
        \label{fig:state_portrait_re8k_E}%
    \end{subfigure}\\
    \vspace{1mm}
    \begin{subfigure}[b]{0.48\textwidth}
        \setlength{\unitlength}{1.0cm}
        \centering
        \begin{picture}(8.0,5.6)
            \put(0.41,0.0){\includegraphics[scale=0.5]{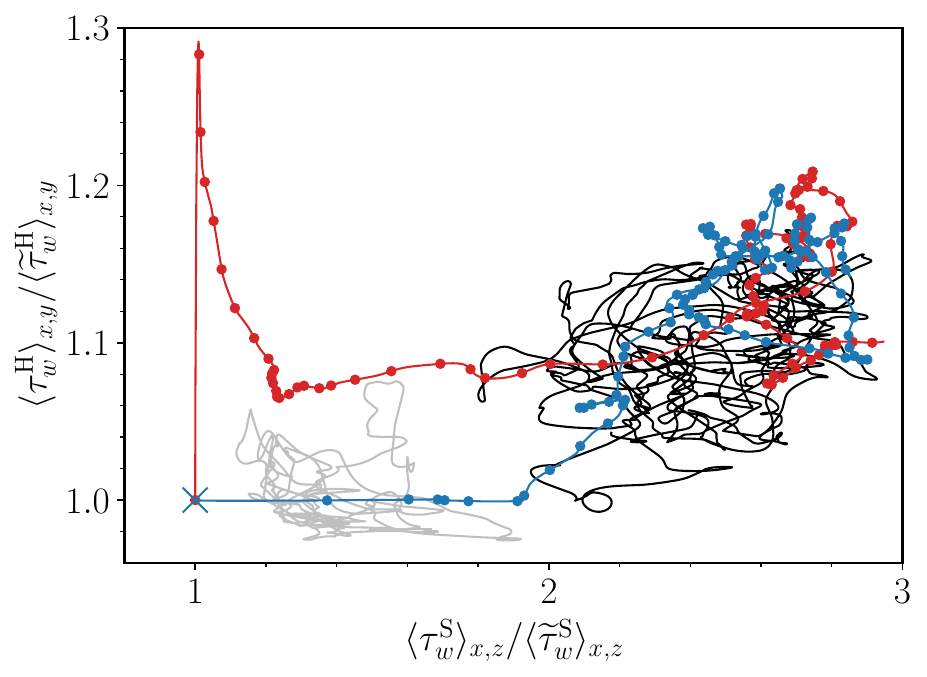}}%
            \put(0.0,0.1){(e)}
        \end{picture}
        \phantomsubcaption
        \label{fig:state_portrait_re5k_tauw}%
    \end{subfigure}          
    \begin{subfigure}[b]{0.48\textwidth}
        \setlength{\unitlength}{1.0cm}
        \centering
        \begin{picture}(8.0,5.6)
            \put(0.32,0.0){\includegraphics[scale=0.5]{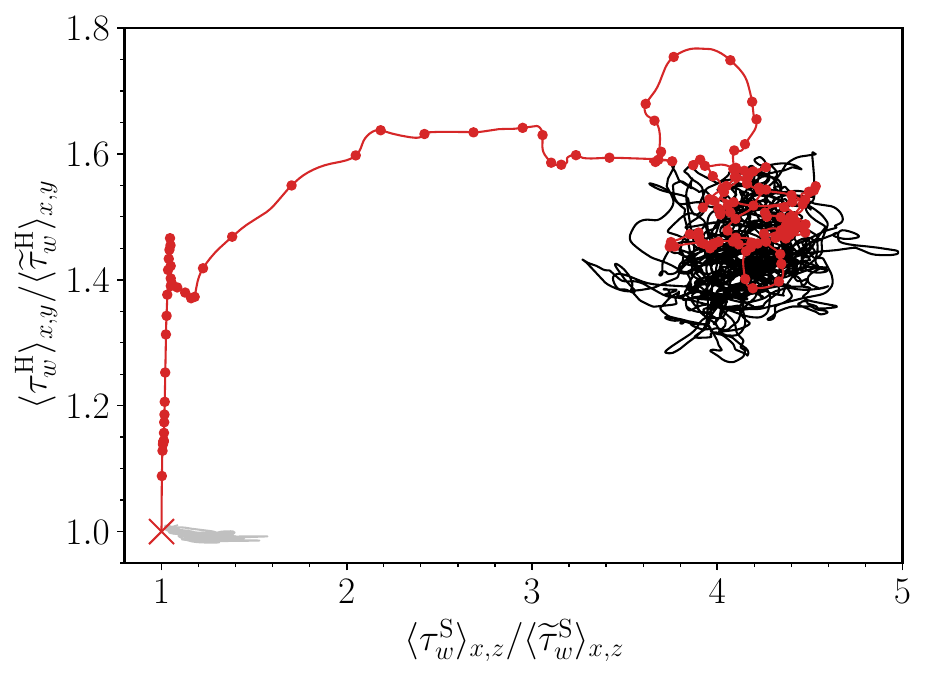}}%
            \put(0.0,0.1){(f)}
        \end{picture}
        \phantomsubcaption
        \label{fig:state_portrait_re8k_tauw}%
    \end{subfigure}      
    \caption{ 
    (\subref{fig:state_portrait_re5k_DI}-\subref{fig:state_portrait_re8k_DI}) $I$ vs.~$D_{\nu}+D_{\mu}$,
    (\subref{fig:state_portrait_re5k_E}-\subref{fig:state_portrait_re8k_E}) $\Ecf$ vs.~$\Estk$, and
    (\subref{fig:state_portrait_re5k_tauw}-\subref{fig:state_portrait_re8k_tauw}) $\langle\tau_w^{\mathrm{H}}\rangle_{x,y}$ vs.~$\langle\tau_w^{\mathrm{S}}\rangle_{x,z}$ for the edge state (gray), turbulence (black), and transition initiated in the Hartmann (red) and the Shercliff (blue) layer.
    The starts of the transition trajectories are marked with a cross, and every subsequent time unit is marked with a filled circle.
    When relevant, the quantities are normalized by their corresponding laminar values, denoted with a tilde $\widetilde{(\cdot)}$.
    Panels (\subref{fig:state_portrait_re5k_DI},\subref{fig:state_portrait_re5k_E},\subref{fig:state_portrait_re5k_tauw}) and (\subref{fig:state_portrait_re8k_DI},\subref{fig:state_portrait_re8k_E},\subref{fig:state_portrait_re8k_tauw}) correspond to case A and case B, respectively. 
    }
    \label{fig:state_portraits}%
\end{figure}

In Fig.~\ref{fig:state_portrait_re5k_DI} the evolution of the total dissipation and the energy input are shown, normalized by their corresponding laminar values. The edge state, plotted in gray, by its definition lies graphically between the laminar and the turbulent attractors.
When transition is initiated in the Hartmann layer for case A (Fig.~\ref{fig:state_portrait_re5k_DI}), it is apparent how the trajectory (red line) is attracted towards the edge state and subsequently gets ejected away towards the turbulent attractor (black line).
A typical sign of an approach to an attractor is the slowing down of the dynamics. To identify slower stages of the transition process, filled circles separated by one advective time unit are plotted.
In Fig.~\ref{fig:state_portrait_re5k_DI}, a cluster of points is clearly visible for the transition trajectory in the vicinity of the edge state.
When transition is instead initiated in the Shercliff layer (blue line), a similar approach towards the edge state albeit less pronounced, is visible after a brief transient.

Other common state space measures are the volumetric streak and cross-flow energy, defined as
\begin{subequations}
  \label{eq:estk_ecf}%
  \begin{gather}
    \Estk = \frac{2}{L}\int_{\Omega} (v_x^{\mathrm{stk}})^2\,\dd \upsilon, \qquad
    \Ecf = \frac{2}{L}\int_{\Omega} \left(v_y^2 + v_z^2\right)\,\dd \upsilon,
    \tag{\theequation a,b}
  \end{gather}
\end{subequations}
respectively.
Since the initial disturbance involves only cross-stream components (see Appendix \ref{sec:synthpert}), the streak energy is initially zero, but as the flow relaxes from this state, energy is transferred into the streamwise perturbation component as well.
In Fig.~\ref{fig:state_portrait_re5k_E}, the approach of both transition trajectories to the edge state, accompanied by the aforementioned slowing down is clearly visible. 

We also investigate the transition process in view of the time-dependent wall-averaged shear stresses,
$\langle\tau_w^{\mathrm{H}}\rangle_{x,y}=\langle \partial v_x/\partial z \rangle_{x,y}|_{z=1}$ 
(and a similar expression for $\langle\tau_w^{\mathrm{S}}\rangle_{x,z}$).
The laminar corresponding values of the base state are 
$\langle\widetilde{\tau}_w^{\mathrm{H}}\rangle_{x,y} = 20.1858$ and 
$\langle\widetilde{\tau}_w^{\mathrm{S}}\rangle_{x,z} = 5.9090$,
which illustrates the large difference in shear stress between the two boundary layer types.
Fig.~\ref{fig:state_portrait_re5k_tauw} shows that both perturbations induce an immediate upshot in the shear stress on their corresponding walls.
Again, since the initial perturbation of choice (see \eqref{eq:pert_components}) does not contain a streamwise component, both normalized transition trajectories start at the same point in this this state space representation.
However, as the transition migrates into the Shercliff layer, the shear stress on the Hartmann wall is bound to come down while that on the Shercliff wall continues to rise until the turbulent attractor is reached.
Interestingly, the edge state appears capable of reducing the shear stress on the Hartmann wall below its laminar value.

Upon increasing the Reynolds number to $R=400$ (case B), turbulence occupies both the Shercliff and the Hartmann wall as earlier observed in Fig.~\ref{fig:std_turb_re8k}.
In the state space projection $(\Estk,\Ecf)$ (Fig.~\ref{fig:state_portrait_re8k_E}), this manifests itself as a shift of the turbulent attractor away from the laminar fixed point, which remains fixed at the origin.
It is remarkable how close the edge state is to the laminar flow in terms of $\Ecf$, which suggests that the basin of attraction of the laminar flow has shrunken relative to case A.
This implies that considerably weaker perturbations now suffice to induce transition, as verified by comparing the red trajectories in Figs.~\ref{fig:state_portrait_re5k_E} and \ref{fig:state_portrait_re8k_E}.
Although the behavior is less apparent than for the lower Reynolds number, evidence that the transition process initiated in the Hartmann layer is guided by the edge state also in this case, may be found in the close-up inset of Fig.~\ref{fig:state_portrait_re8k_DI}, as well as by the clustering of points around the values of the abscissa occupied by the edge state in Figs.~\ref{fig:state_portrait_re8k_E} and \ref{fig:state_portrait_re8k_tauw}.
These results thus support favorably the notion put forth in Ref.~\cite{khapko2016edge} that the edge states act as mediators of turbulent flow.

%=============================================================================
\section{Conclusions and outlook}
\label{sec:conclusions_outlook}%

\begin{figure}
    \centering
    \begin{subfigure}[b]{0.46\textwidth}
        \setlength{\unitlength}{1.0cm}
        \centering
        \begin{picture}(6.5,5.1)
            \put(0.0,0.3){\includegraphics[scale=0.5]{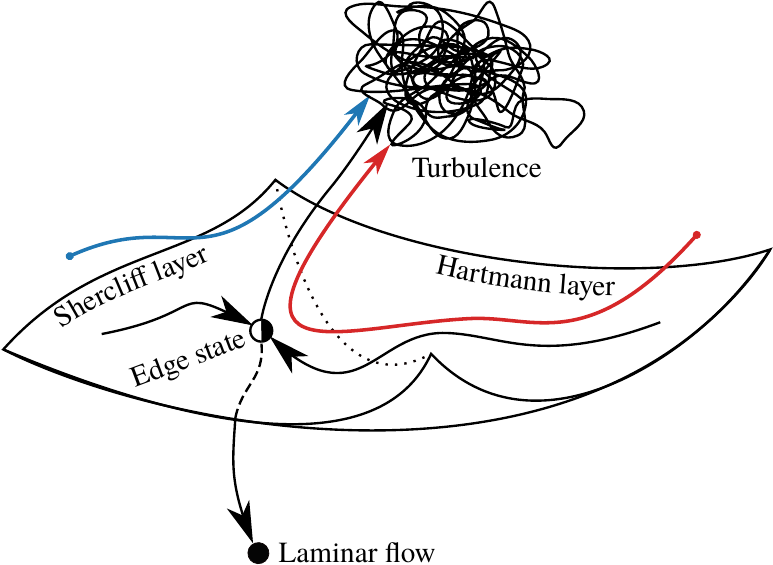}}%
            \put(-0.15,0.1){(a)}
        \end{picture}       
        \phantomsubcaption
        \label{fig:sketch_state_space}%
    \end{subfigure}          
    \begin{subfigure}[b]{0.46\textwidth}
        \setlength{\unitlength}{1.0cm}
        \centering
        \begin{picture}(7.0,5.1)
            \put(0.0,-0.05){\includegraphics[scale=0.5]{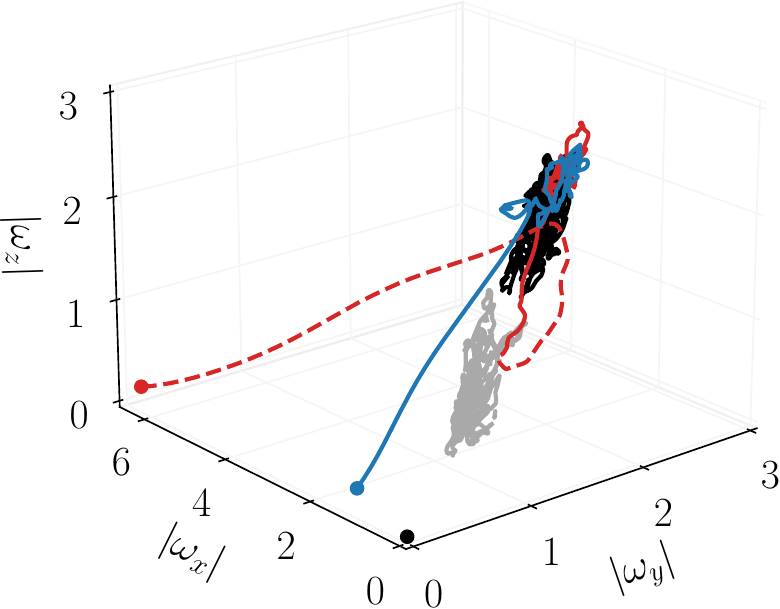}}%
            \put(-0.15,0.1){(b)}
        \end{picture}
        \phantomsubcaption
        \label{fig:3d_state_space}%
    \end{subfigure}
    \caption{(\subref{fig:sketch_state_space}) The classical state space sketch illustrating the edge manifold acting as a basin boundary of the laminar attractor with the edge state plotted as a single saddle state. States on the manifold whose \eg~center of enstrophy are located on the Hartmann ($\yh<\zh$) and the Shercliff ($\yh\geq\zh$) wall are delimited by a dotted line. 
    (\subref{fig:3d_state_space}) Verification of the preceding sketch using perturbation vorticity data from the MHD duct simulations of case A.
    Segments of the trajectories for which $\yh\geq\zh$ ($\yh<\zh$) are shown with a solid (dashed) line.
    The edge state (gray) guides transition to turbulence (black) initiated in both the Hartmann (red) and the Shercliff (blue) layer. The base laminar state is marked with a black circle.}
    \label{fig:state_space}%
\end{figure}

Whereas earlier works on MHD duct flows have exclusively focused on flow laminarization \cite{branover_1978,krasnov2013patterned}, the forward transition scenario is here investigated for the first time from a fully nonlinear point of view. In this computational study we have demonstrated, for a representative set of parameter values and under a symmetry restriction, that transition to turbulence in MHD duct flow occurs by the finite-amplitude destabilization of the Shercliff layer. Conversely, as opposed to the picture advanced by early experimental studies (see \cite{branover_1978} for a review), no transition route involving the Hartmann layers has been observed. Computations of edge states for the same parameters indicate, consistently with the description of the transition process, similar spatial localization next to the Shercliff walls. 
The existence of these nonlinear recurrent solutions, together with their localization properties, suggests that actual transition depends on a state space scaffold built around specific solutions, all localized spatially within the Shercliff layer. 
This crucial role of the Shercliff layer is not the outcome of a competition between different solutions with different localization properties and stability properties:
only nonlinear solutions localized in the Shercliff layer have been found, whereas no genuine solution has been detected on the Hartmann walls. While such solutions cannot be strictly excluded, it is likely that their stability properties exclude them as edge states.
Inspired by the original sketches in \eg~\cite{toh2003periodic,duguet2008transition}, a graphical summary of the above conclusions is given in Fig.~\ref{fig:sketch_state_space}.
For simplicity, we have here plotted the edge state as a single saddle point.
With reference to Fig.~\ref{fig:enstrophy_centroid}, initial conditions $\vv^{\mathrm{bis}}$ for which $\yh<\zh$, are separated from those with $\yh\geq\zh$ by a dotted line.
Further support for this conceptual picture is provided in the three-dimensional state portrait in Fig.~\ref{fig:3d_state_space} constructed using perturbation vorticity components (whose norms are evaluated using \eqref{eq:jx_norm}) extracted from the non-linear DNS.

\begin{figure}
    \centering
    \begin{subfigure}[b]{0.40\textwidth}
        \setlength{\unitlength}{1.0cm}
        \centering
        \begin{picture}(6.0,5.6)
            \put(-0.15,0.05){\includegraphics[scale=0.5]{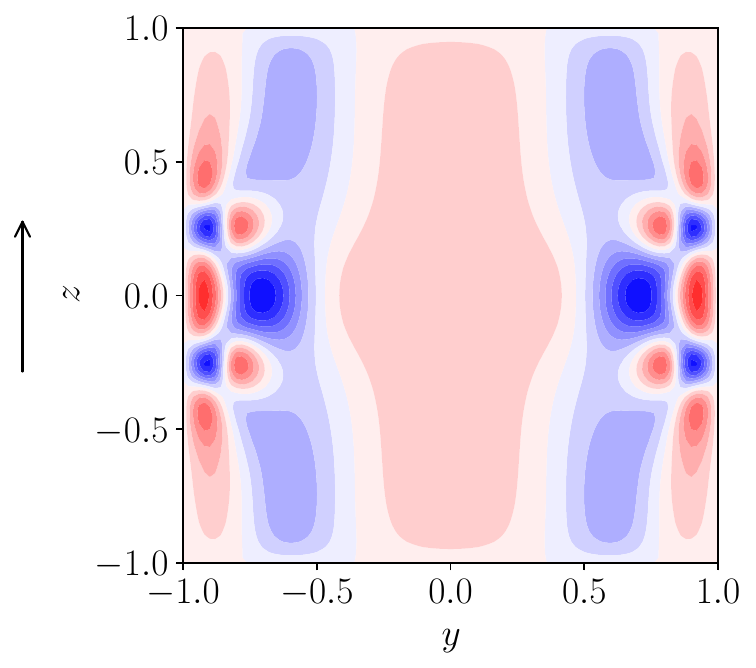}}%
            \put(-0.15,0.0){(a)}
        \end{picture}       
        \phantomsubcaption
        \label{fig:edge_state2d}%
    \end{subfigure}          
    \begin{subfigure}[b]{0.44\textwidth}
        \setlength{\unitlength}{1.0cm}
        \centering
        \begin{picture}(7.5,5.6)
            \put(-0.15,0.05){\includegraphics[scale=0.5]{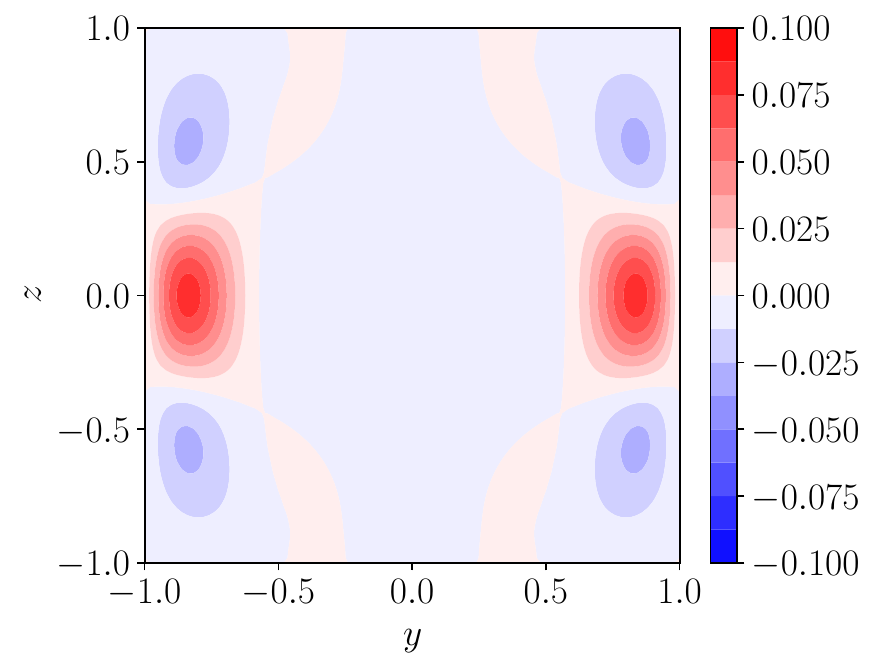}}%
            \put(-0.15,0.0){(b)}
        \end{picture} 
        \phantomsubcaption
        \label{fig:energy_mode}%
    \end{subfigure}
    \caption{Comparison of (\subref{fig:edge_state2d}) the streak velocity $\langle v_x^{\mathrm{stk}}\rangle_x$ of the edge state for case B at time $t=1\,144.8$, and (\subref{fig:energy_mode}) the streamwise velocity component of the least energy stable $x$-independent mode in a square duct at $\Ha=20$ \cite{boeck_etal_2024}.
    The black arrow in (\subref{fig:edge_state2d}) shows the diretion of the magnetic field, and for visualization purposes, 
    the edge state in the quarter duct is extended to the full duct.
    The amplitude of the energy stability mode is normalized to have the same maximum as the streaks of the edge state.
    }
    \label{fig:edge_vs_energy}%
\end{figure}

The dynamics exhibited by the edge states are truly three-dimensional as can be observed in Fig.~\ref{fig:edgestate_re5k}--\ref{fig:std_edge}, especially Fig.~\ref{fig:edgestate_re5k_intense}. Nevertheless, it is instructive to compare and contrast them with the spatial structure of the solutions that emerge from the various analysis methods developed in the stability literature.
In order of ascending $\Rey$, the first critical value of interest is the energy Reynolds number $\Rey_E$, below which no transient energy growth is possible for any perturbation to the base flow. It was computed recently for the MHD square duct in \cite{boeck_etal_2024}. 
For $\Ha=20$ the corresponding value is $\Rey_E= 260$ with $k_x\approx 4$.
In Fig.~\ref{fig:energy_mode}, the least energy-stable mode with $k_x=0$ is shown (as a side note, we mention that unlike in HD square ducts \cite{biau2008transition}, modes with $k_x=0$ are suboptimal with respect to transient growth in the MHD case \cite{krasnov2010optimal,boeck_etal_2024}). A staggered pattern of streaks, with low wavenumber, occupies the Shercliff layer only. Above $Re_E$, linear transient growth of perturbations is possible. The coherent structures associated with the largest energy amplification were computed numerically by \cite{krasnov2010optimal}, \cf~their figure 7c for the optimal $k_x=0$ modes. Similar streak patterns occur, again confined to the Shercliff layer. The nonlinear generalization of these optimal structures, the minimal seeds \cite{kerswell2014optimization}, have not yet been computed for the three-dimensional MHD case (see \cite{camobreco2020subcritical} for quasi-two-dimensional computations). All these streak patterns, with their different wavenumbers reflecting different values of $\Rey_{\tau}$, are directly comparable
to the instantaneous streak pattern of the present (nonlinear) edge states. As an illustration, the instantaneous $x$-averaged streamwise velocity of the edge state computed for case B in Fig.~\ref{fig:edge_state2d} features the staggered pattern of low- and high-speed streaks typically observed in edge states with bursting dynamics \cite{kreilos2013edge}. Due to the presence of the corners, the structures are spanwise localized.
The comparison in Fig.~\ref{fig:edge_vs_energy} corroborates the central role of the Shercliff layers in supporting basic perturbation growth mechanisms such as the \emph{lift-up effect} \cite{landahl1980note}. This mechanism is one key element to the self-sustaining processes \cite{hamilton1995regeneration} (of which the edge state is a manifestation). These mutual links justify why the very same Shercliff region morphs at higher values or $\Rey$ (or $R$) into a promoter of transition at the expense of the inactive Hartmann layers and bulk region.
Although the bisection algorithm does not identify any unstable exact solutions localized in the Hartmann layer, we cannot exclude their existence. Such structures have for instance been identified in \eg~energy stability analysis \cite{boeck_etal_2024}, but are there found to be associated with a relatively larger $Re_E$. 
Likewise, they would be expected to be unimportant for transition to and sustenance of turbulence.

With the edge state localized on the Shercliff wall, the situation is locally similar to the case of a channel subject to spanwise magnetic field.
In that configuration, we recently showed that the edge state changes from a recurrent cycle of calm and bursting stages to spatio-temporal chaos as the interaction parameter $\Nc$ is increased \cite{brynjell-rahkola_etal_2024}.
A similar change in dynamics is observed also in this case in passing from $\Nc=0.05$ (case B) to $0.08$ (case A). To what extent the verification of this trend is fortuitous and depends on other parameters is presently unknown. As for the MHD channel, the turbulent and the edge regime appear closer in state space for $\Nc=0.08$ than for $0.05$ (see Fig.~\ref{fig:state_portraits}).
This is interpreted in Ref.~\cite{brynjell-rahkola_etal_2024} as the approach towards a tipping point as $\Nc$ is increased. This event would correspond to a generalized saddle node bifurcation, beyond which the only attracting solution is the laminar one.
Such a result would be consistent with the known difficulty of both initiating and maintaining turbulence at high $\Nc$.

Eventually, one may speculate how the results obtained with the present symmetric model generalize to the full duct.
Since the spatial localization of the edge states to the Shercliff wall is commensurate with the preferred region of transient growth, no significant change in the edge state nor the transition route is expected.
In fact, it is shown in Appendix \ref{sec:turb_stat} that the turbulence activity on the Hartmann (Shercliff) walls is lower (higher) in the full duct than in the quarter duct.
This intuitively suggests a possibly even stronger preference for the Shercliff walls in the full duct than what is reported here.

Although the drag in a MHD duct, due to the very steep velocity gradients, is dominated by the Hartmann layers, the state portraits in Figs.~\ref{fig:state_portrait_re5k_tauw} and \ref{fig:state_portrait_re8k_tauw} reveal that the edge state for both investigated parameter configurations at least momentarily entails lower values of $\langle\tau_w^{\mathrm{H}}\rangle_{x,y}$ than the laminar flow.
This suggests a novel control strategy for reducing pumping requirements in liquid metal applications, \ie~by maintaining the flow on the edge manifold.
Given that there are an infinite number of unstable periodic orbits embedded in a strange attractor, control strategies have been proposed to turbulent flow \cite{kawahara2005laminarization}. However, in light of the above observation of wall shear stress, the objective need here not be a full laminarization of the flow, but a mere restriction of it to the edge state. Such an approach might potentially be less demanding since it can suffice to stabilize a periodic orbit \cite{pyragas1992continuous}.

The present results are expected to hold for larger finite aspect ratios, as demonstrated in Ref.~\cite{krasnov2015patterned}. The connection between the limiting configuration of a Hartmann channel and finite aspect ratio ducts is still an open problem, although three-dimensional edge states in the Hartmann channel have been recently reported in \cite{shi2023effect}. 
It is of interest to consider also the influence of the Hartmann number on the reported transition routes. Unlike the Reynolds number, the $\Ha$-number directly influences the base state through \eqref{eq:baseflow_eq}. The choice of boundary conditions can also strongly influence the flow. For instance if the sidewalls are electrically conducting (\ie~Hunt's flow) the base flow develops strong wall jets, likely to be more unstable, which makes the transition scenario fully different \cite{kinet2009instabilities,priede2010linear}.
One may also study the effect of changing the orientation of the magnetic field.
For instance, in the case of a streamwise magnetic field \cite{dong2012secondary}, the equivalence of all walls as in ordinary HD is recovered.
In this case the streaks become inconspicuous to the Lorentz force.
Such parameter variations will be left for future work.

%=============================================================================
\section*{Acknowledgments}
M.B.R.~and T.B.~acknowledge financial support from The German Research Foundation (DFG) under project no.~470628784. 
The authors gratefully acknowledge the Gauss Centre for Supercomputing e.V.~(\url{www.gauss-centre.eu}) for funding this project by providing computing time (project no.~\textit{pn49su}) on the GCS Supercomputer SuperMUC-NG at Leibniz Supercomputing Centre (\url{www.lrz.de}). Post-processing and data visualization were performed at the local computing center of TU Ilmenau.
Y.D.~thanks TU Ilmenau for the hospitality during his research visits related to this project.

\appendix

%=============================================================================
\section{Numerical resolution}
\label{sec:resolutions}%

In order to assess the spatial resolution of the DNS simulations and ensure that all length scales of the flow are properly resolved, inner units as defined in \S\ref{sec:length_domain} are considered.
In Table \ref{tab:retau} the friction Reynolds numbers $\Rey_{\tau}^{\mathrm{S}}$ and $\Rey_{\tau}^{\mathrm{H}}$ evaluated on the symmetry lines $(y,z=0)$ and $(y=0,z)$, as well as on the midlines $(y,z=0.5)$ and $(y=0.5,z)$ of the quarter duct are presented.
For comparison, the wall averaged quantities
\begin{subequations}
  \label{eq:retau_aver}%
  \begin{gather}
    \langle\Rey_{\tau}^{\mathrm{H}}\rangle = \left(\Rey\,\overline{\left\langle\left.\frac{\partial v_x}{\partial z}\right|_{z=1}\right\rangle_{x,y}}\right)^\frac{1}{2}, \qquad
    \langle\Rey_{\tau}^{\mathrm{S}}\rangle = \left(\Rey\,\overline{\left\langle\left.\frac{\partial v_x}{\partial y}\right|_{y=1}\right\rangle_{x,z}}\right)^\frac{1}{2}
    \tag{\theequation a,b}
  \end{gather}
\end{subequations}
are shown as well.
For both the full and the quarter duct, the friction Reynolds number reaches its largest values on the symmetry lines. 
Numerically resolving these regions thus imposes the strongest restrictions on the computational grid.

\begin{table}[!t]
    \caption{
    Comparison of the friction Reynolds numbers on the Hartmann and the Shercliff walls between different simulation cases.
    The Reynolds numbers are evaluated at the two symmetry lines, at the two midlines of the quarter duct, and averaged over the two walls.
    For the full duct \cite{brynjell-rahkola_2024}, the data is symmetrized and averaged between the two pairs of walls.
    }
    \label{tab:retau}%
    \begin{ruledtabular}
        \begin{tabular}{lcccccccc}
            Geometry & $L$ & $\Rey$ & $\left.\Rey_{\tau}^{\mathrm{H}}\right|_{y=0}$ & $\left.\Rey_{\tau}^{\mathrm{H}}\right|_{y=0.5}$ & $\langle\Rey_{\tau}^{\mathrm{H}}\rangle$
            & $\left.\Rey_{\tau}^{\mathrm{S}}\right|_{z=0}$ & $\left.\Rey_{\tau}^{\mathrm{S}}\right|_{z=0.5}$ & $\langle\Rey_{\tau}^{\mathrm{S}}\rangle$ \\[1ex]
            \colrule
            Full duct & $8\pi$ & 5000 & 359.9 & 340.3 & 329.9 & 281.6 & 281.6 & 266.1 \\
            Quarter duct & $8\pi$ & 5000 & 375.4 & 326.6 & 332.7 & 320.0 & 293.0 & 267.4 \\
            Quarter duct & $2\pi$ & 5000 & 371.9 & 324.1 & 327.6 & 324.7 & 288.1 & 262.2 \\
            Quarter duct & $\pi$ & 8000 & 588.9 & 490.3 & 475.6 & 560.8 & 454.7 & 435.3
        \end{tabular}
    \end{ruledtabular}
\end{table}

The numerical resolution used in the different simulations discussed in \S\ref{sec:edge_state} and in Appendix \ref{sec:turb_stat} are given in Table \ref{tab:mesh}. 
Following \cite{krasnov2012numerical}, the elements are defined in $y$ and $z$ as a blend between the Gauss-Lobatto-Chebyshev (GLC) discretization and uniform spacing, such that
\begin{subequations}
  \label{eq:mesh_stretch}%
  \begin{gather}
    z(i) = \gamma_z z_{\mathrm{GLC}}(i) + (1-\gamma_z) z_{\mathrm{eq}}(i), \quad 
    z_{\mathrm{GLC}}(i) = \sin\left(\frac{\pi i}{n_{\mathrm{el},z}} - \frac{\pi}{2}\right), \quad
    z_{\mathrm{eq}}(i) = \frac{2i}{n_{\mathrm{el},z}} - 1,
    \tag{\theequation a,b,c}
  \end{gather}
\end{subequations}
with $i\in[0,n_{\mathrm{el},z}]$ for $z$, and similarly for $y$.
The streamwise $x$-extent of the domain is discretized using $n_{\mathrm{el},x}$ equidistant elements.

In the simulations of wall turbulence, the elements are smallest next to the walls and they smoothly stretch when approaching the symmetry lines.
With reference to eq.~\eqref{eq:mesh_stretch}, the total element count reported in the third column of Table \ref{tab:mesh} corresponds to $n_{\mathrm{el}}=n_{\mathrm{el},x}\times (n_{\mathrm{el},y}/2)\times (n_{\mathrm{el},z}/2)$.
The mesh parameters $n_{\mathrm{el},z}$, $\gamma_z$, $n_{\mathrm{el},y}$, $\gamma_y$ and $n_{\mathrm{el},x}$ are carefully chosen to obtain resolutions similar or better than previous studies on turbulent flow in HD \cite{uhlmann2007marginally,ohlsson2010direct,vinuesa2014aspect,atzori2021intense} and MHD \cite{krasnov2012numerical} duct geometries. 
Note however that since the figures in Table \ref{tab:mesh} are based on values at the symmetry lines where the friction Reynolds numbers attain their maxima, the resolution is mostly better than what is reported.
(For details of the resolution in the full duct used as reference in Appendix \ref{sec:turb_stat}, the reader is referred to \cite{brynjell-rahkola_2024}.)

Following \cite{brynjell-rahkola_etal_2024}, bisection (see \S\ref{sec:edge_state}) is done on the same mesh topology as the turbulence simulations but with a lower polynomial order $N$.
This is justified since edge states and other lower branch states feature much less energy in the small spatial scales compared to developed turbulence.

An accurate capture of the transition process requires refinements in the mesh topology.
In order to induce transition (see \S\ref{sec:transition}), a synthetic perturbation in the form of a localized vortex pair \cite{chevalier_etal_2007} is imposed on the laminar solution (see Appendix \ref{sec:synthpert}).
To resolve the strong updraft of low-speed fluid along the symmetry lines that this perturbation generates, the elements next to the symmetry lines have the same size as the ones next to the walls.
Thus, the element distributions given by eq.~\eqref{eq:mesh_stretch} are linearly transformed \ie~$z'=(z+1)/2$ (and similar for $y$).
In this case, the element count in Table \ref{tab:mesh} is given by $n_{\mathrm{el}}=n_{\mathrm{el},x}\times n_{\mathrm{el},y}\times n_{\mathrm{el},z}$. 

\begin{table}[!t]
    \caption{
    Mesh parameters for the different simulation cases considered.
    $n_{\mathrm{el}}$ and $N$ denote the number of spectral elements in the mesh and the polynomial order, respectively.
    In the following four columns, the grid spacing between neighboring mesh points in the different directions are reported.
    The first and the second entry in each tuple refer to the smallest and the largest node spacing in the mesh, respectively.
    The two last columns report the width of the SE next to the walls, \ie~the distance of the $N$th point off the wall:
    $y_{N+}^{\mathrm{S}} = y(n_{\mathrm{el},y})_+^{\mathrm{S}}-y(n_{\mathrm{el},y}-1)_+^{\mathrm{S}}$ and 
    $z_{N+}^{\mathrm{H}} = z(n_{\mathrm{el},z})_+^{\mathrm{H}}-z(n_{\mathrm{el},z}-1)_+^{\mathrm{H}}$.}
    \label{tab:mesh}%
    \begin{ruledtabular}
        \begin{tabular}{lccccccccccc}
            Type & $L$ & $\Rey$ & $n_{\mathrm{el}}$ & $N$ & $\Delta x_+^{\mathrm{H}}$ & $\Delta x_+^{\mathrm{S}}$ & $\Delta y_+^{\mathrm{S}}$ & $\Delta z_+^{\mathrm{H}}$ & $y_{N+}^{\mathrm{S}}$ & $z_{N+}^{\mathrm{H}}$ \\[1ex]
            \colrule
            Turbulence & $8\pi$ & 5000 & 88\,540 & 11 & (1.116, 5.530) & (0.951, 4.714) & (0.091, 3.424) & (0.084, 3.866) & 3.311 & 3.044 \\
            Turbulence & $2\pi$ & 5000 & 36\,000 & 11 & (1.073, 5.319) & (0.937, 4.643) & (0.072, 2.741) & (0.081, 3.009) & 2.633 & 2.936 \\
            Turbulence & $\pi$ & 8000 & 55\,930 & 11 & (1.085, 5.375) & (1.033, 5.118) & (0.065, 3.396) & (0.066, 3.463) & 2.337 & 2.388 \\[0.5ex]
            Bisection & $2\pi$ & 5000 & 36\,000 & 7 & (2.498, 8.152) & (2.180, 7.116) & (0.169, 4.201) & (0.188, 4.612) & 2.633 & 2.936 \\      
            Bisection & $\pi$ & 8000 & 55\,930 & 7 & (2.524, 8.239) & (2.404, 7.845) & (0.150, 5.205) & (0.153, 5.308) & 2.337 & 2.388 \\[0.5ex]   
            Transition & $2\pi$ & 5000 & 316\,368 & 11 & (0.550, 2.728) & (0.480, 2.381) & (0.053, 1.207) & (0.055, 1.397) & 1.910 & 2.011 \\  
            Transition & $\pi$ & 8000 & 235\,200 & 11 & (0.680, 3.369) & (0.647, 3.208) & (0.088, 1.922) & (0.068, 2.091) & 3.202 & 2.475
        \end{tabular}
    \end{ruledtabular}
\end{table}

%=============================================================================
\section{Initial perturbation}
\label{sec:synthpert}%

The initial perturbation used to induce transition corresponds to a localized vortex pair documented in \cite{chevalier_etal_2007} and its definition is repeated here for completeness.
In order to position the perturbation on different walls, we define it in a local Cartesian coordinate system $(x_{\mathrm{p}},y_{\mathrm{p}},z_{\mathrm{p}})$ related to the global coordinate system $(x,y,z)$ by a translation and a rotation
\begin{equation}
    \label{eq:pert_transform}%
    \begin{pmatrix}
        x_{\mathrm{p}} \\
        y_{\mathrm{p}} \\
        z_{\mathrm{p}}
    \end{pmatrix} =
    \begin{bmatrix}
        1 & 0 & 0 \\
        0 & \cos(\alpha) & \sin(\alpha) \\
        0 & -\sin(\alpha) & \cos(\alpha)
    \end{bmatrix}
    \begin{pmatrix}
        x-x_{\mathrm{c}} \\
        y-y_{\mathrm{c}} \\
        z-z_{\mathrm{c}}
    \end{pmatrix}.
\end{equation}
To yield an appropriately sized perturbation, these local coordinates are subsequently scaled as $\xh=x_{\mathrm{p}}/x_{\mathrm{s}}$, $\yh=y_{\mathrm{p}}/y_{\mathrm{s}}$ and $\zh=z_{\mathrm{p}}/z_{\mathrm{s}}$, where $x_{\mathrm{s}}$, $y_{\mathrm{s}}$ and $z_{\mathrm{s}}$ are suitable scale factors.
Next, the following streamfunction is introduced
\begin{subequations}
  \label{eq:pert_streamfunction}%
  \begin{gather}
    \Psi(\xh,\yh,\zh) = -(A z_{\mathrm{s}})f(\xh,\zh)g(\yh), \qquad
    \mathcal{U} = 0, \qquad
    \mathcal{V} = \frac{\partial \Psi}{\partial z_{\mathrm{p}}}, \qquad
    \mathcal{W} = -\frac{\partial \Psi}{\partial y_{\mathrm{p}}},
    \tag{\theequation a,b,c,d}
  \end{gather}
\end{subequations}
where
\begin{subequations}
  \label{eq:pert_functions}%
  \begin{gather}
    f(\xh,\zh) = \xh\zh \exp(-\xh^2 - \zh^2), \qquad
    g(\yh) = \yh^3\exp(-\yh^2),
    \tag{\theequation a,b}
  \end{gather}
\end{subequations}
and $A$ is the amplitude of the perturbation.
The different velocity components of the perturbation are given by
\begin{subequations}
  \label{eq:pert_components}%
  \begin{gather}
    \mathcal{U} = 0, \qquad
    \mathcal{V} = -A\xh g(\yh) \exp(-\xh^2 - \zh^2) (1-2\zh^2), \qquad
    \mathcal{W} = A\frac{z_{\mathrm{s}}}{y_{\mathrm{s}}}\xh\zh\frac{\dd g}{\dd \yh}\exp(-\xh^2 - \zh^2).
    \tag{\theequation a,b,c}
  \end{gather}
\end{subequations}
The perturbation components \eqref{eq:pert_components} are finally transformed back to the global velocity components $(v_x,v_y,v_z)$ via the transpose of the rotation matrix in \eqref{eq:pert_transform}.

%=============================================================================
\section{Turbulence statistics}
\label{sec:turb_stat}%

It is interesting to compare quantitatively the statistics of the turbulent flow in the presence, or not, of the quarter duct symmetries, although nothing guarantees that these statistics should match. To this end, the toolbox documented in \cite{vinuesa_etal_2017} is used and flow statistics are gathered during approximately 20 flow-throughs (\ie~around 520 advective time units) for a case with parameters $\Rey=5000$, $\Ha=20$ and $L=8\pi$. As before, these simulations are carried out using \textit{NEK5000} \cite{nek5000}.
Comparing first the friction Reynolds numbers in Table \ref{tab:retau} for the full and the quarter duct, local variations are notable.
This is particularly the case along the symmetry lines, where $\Rey_{\tau}^{\mathrm{S}}$ and $\Rey_{\tau}^{\mathrm{H}}$ in the quarter duct are larger compared to the full duct.
In contrast, the wall-averaged values are in close agreement.

\begin{figure}[!t]
  \centering
  \begin{subfigure}[b]{0.49\textwidth}
    \setlength{\unitlength}{1.0cm}
    \begin{picture}(8.5,4.9)
      \put(0.44,0){\includegraphics[scale=0.5]{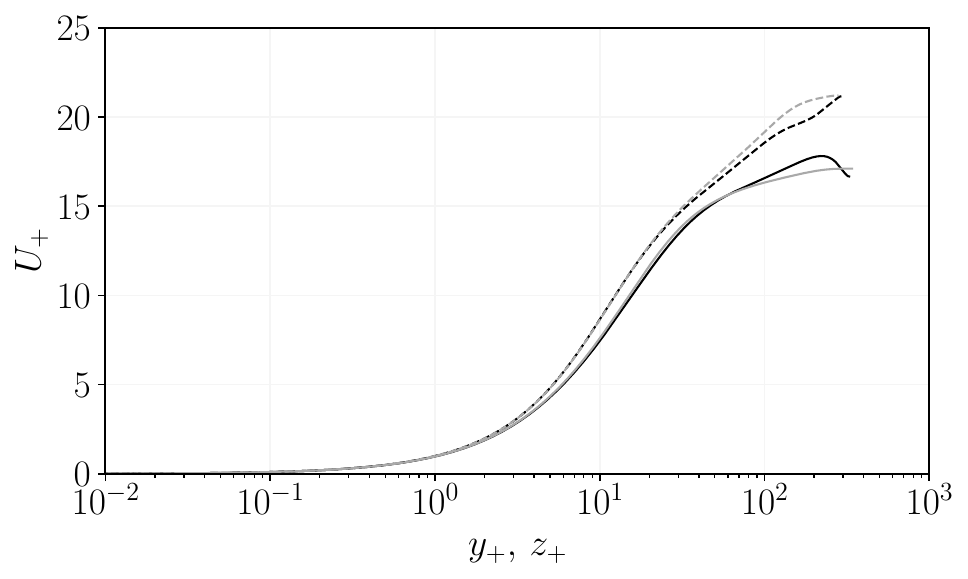}}%
      \put(-0.1,0.0){(a)}
    \end{picture}
    \phantomsubcaption
    \label{fig:stat_Up}%
  \end{subfigure}
  \hfill
  \begin{subfigure}[b]{0.49\textwidth}
    \setlength{\unitlength}{1.0cm}
    \begin{picture}(8.5,4.9)
      \put(0.37,0){\includegraphics[scale=0.5]{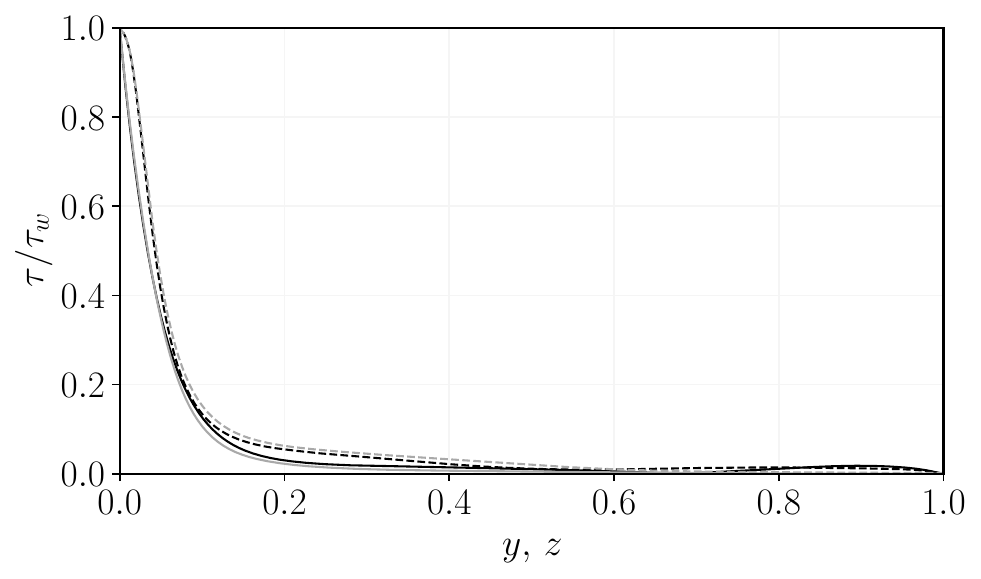}}%
      \put(-0.1,0.0){(b)}
    \end{picture}
    \phantomsubcaption
    \label{fig:stat_tau}%
  \end{subfigure}\\
  \begin{subfigure}[b]{0.49\textwidth}
    \setlength{\unitlength}{1.0cm}
    \begin{picture}(8.5,4.9)
      \put(0.50,0){\includegraphics[scale=0.5]{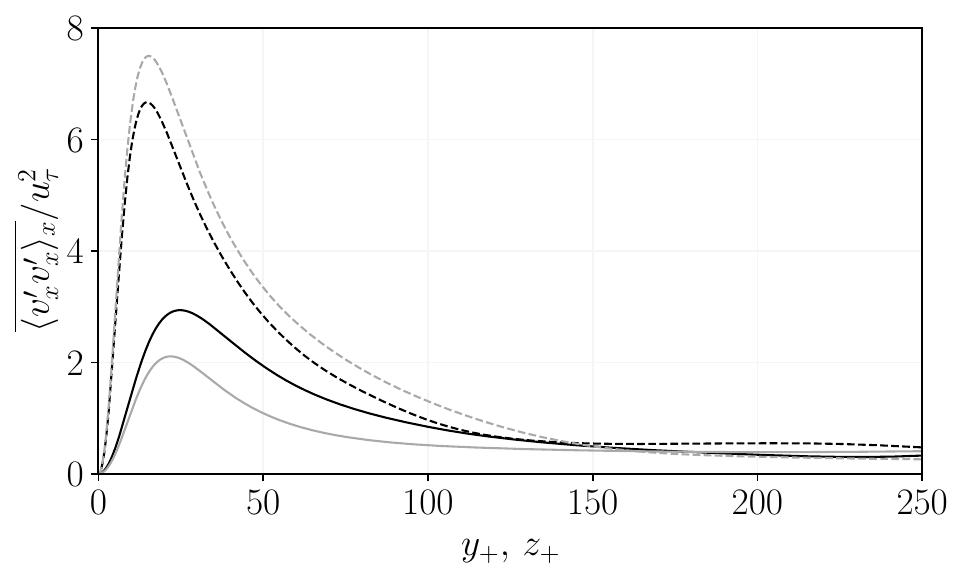}}%
      \put(-0.1,0.0){(c)}
    \end{picture}
    \phantomsubcaption
    \label{fig:stat_uu}%
  \end{subfigure}
  \hfill
  \begin{subfigure}[b]{0.49\textwidth}
    \setlength{\unitlength}{1.0cm}
    \begin{picture}(8.5,4.9)
      \put(0.27,0){\includegraphics[scale=0.5]{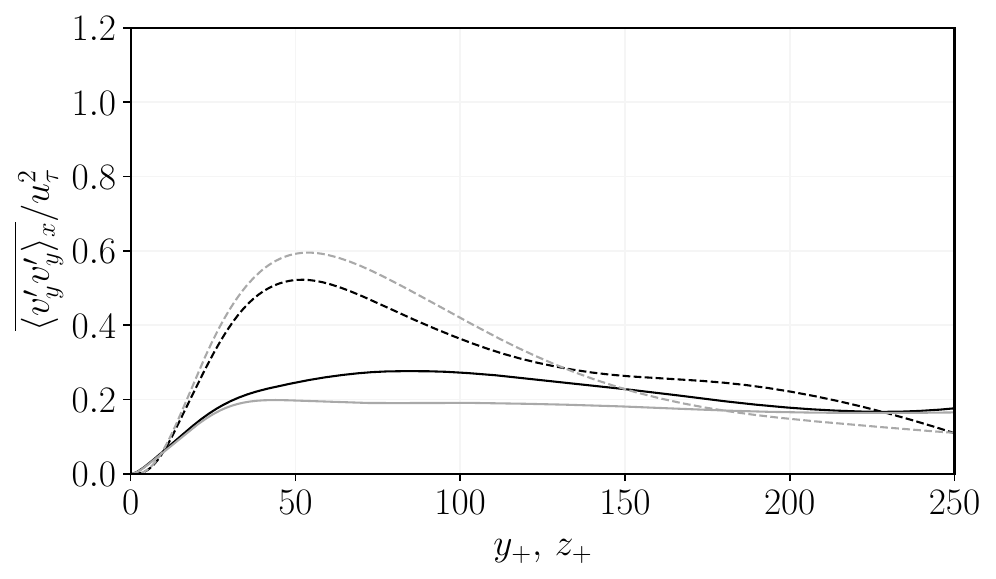}}%
      \put(-0.1,0.0){(d)}
    \end{picture}
    \phantomsubcaption
    \label{fig:stat_vv}%
  \end{subfigure}\\
  \begin{subfigure}[b]{0.49\textwidth}
    \setlength{\unitlength}{1.0cm}
    \begin{picture}(8.5,4.9)
      \put(0.27,0){\includegraphics[scale=0.5]{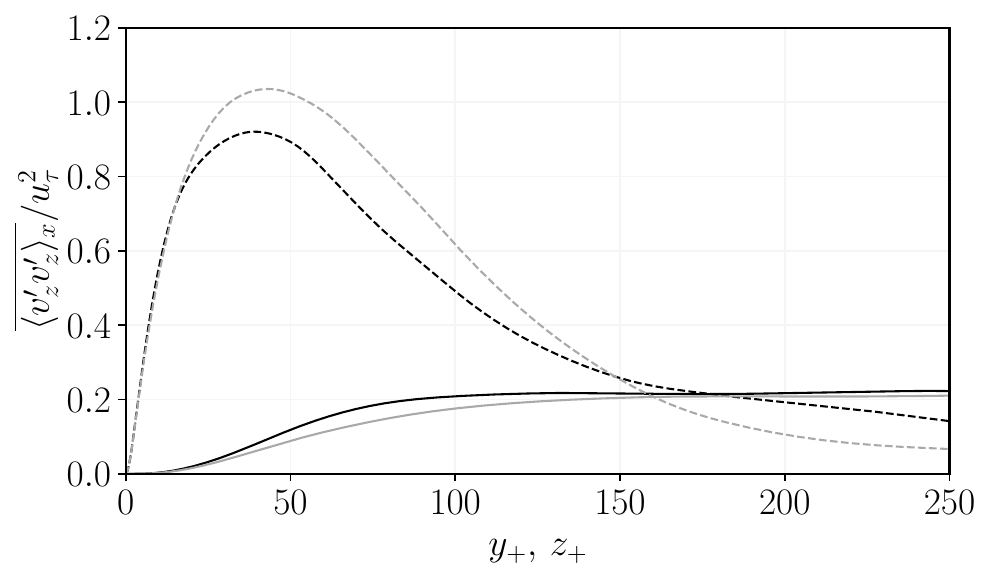}}%
      \put(-0.1,0.0){(e)}
    \end{picture}
    \phantomsubcaption
    \label{fig:stat_ww}%
  \end{subfigure}
  \hfill
  \begin{subfigure}[b]{0.49\textwidth}
    \setlength{\unitlength}{1.0cm}
    \begin{picture}(8.5,4.9)
      \put(0.27,0){\includegraphics[scale=0.5]{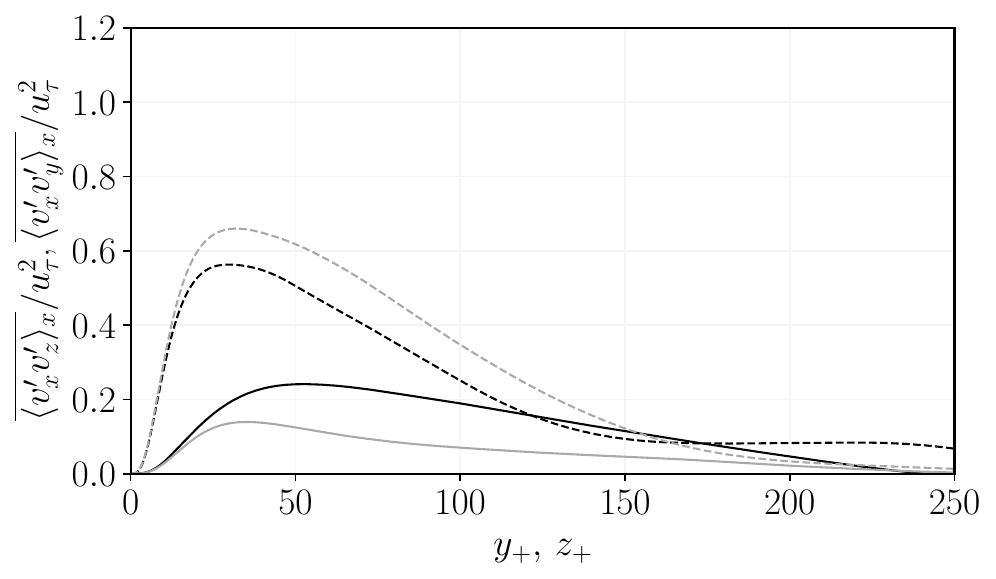}}%
      \put(-0.1,0.0){(f)}
    \end{picture}
    \phantomsubcaption
    \label{fig:stat_uw_uv}%
  \end{subfigure}  
  \caption{Symmetrized turbulence statistics in the full duct \cite{brynjell-rahkola_2024} (gray lines) \emph{versus} turbulence statistics computed in the symmetric quarter duct (black lines).
  $z_+$-profiles across the Hartmann layer for $y=0.5$ plotted with solid lines, $y_+$-profiles across the Shercliff layer for $z=0.5$ plotted with dashed lines.
  (\subref{fig:stat_Up}) Mean velocity $U_+$,
  (\subref{fig:stat_tau}) Mean shear stress, 
  (\subref{fig:stat_uu}--\subref{fig:stat_ww}) Reynolds stresses in the $x$-, $y$- and $z$-directions.
  (\subref{fig:stat_uw_uv}) Reynolds shear stresses in the $xz$- (solid) and the $xy$-direction (dashed).}
  \label{fig:turb_stat}%
\end{figure}
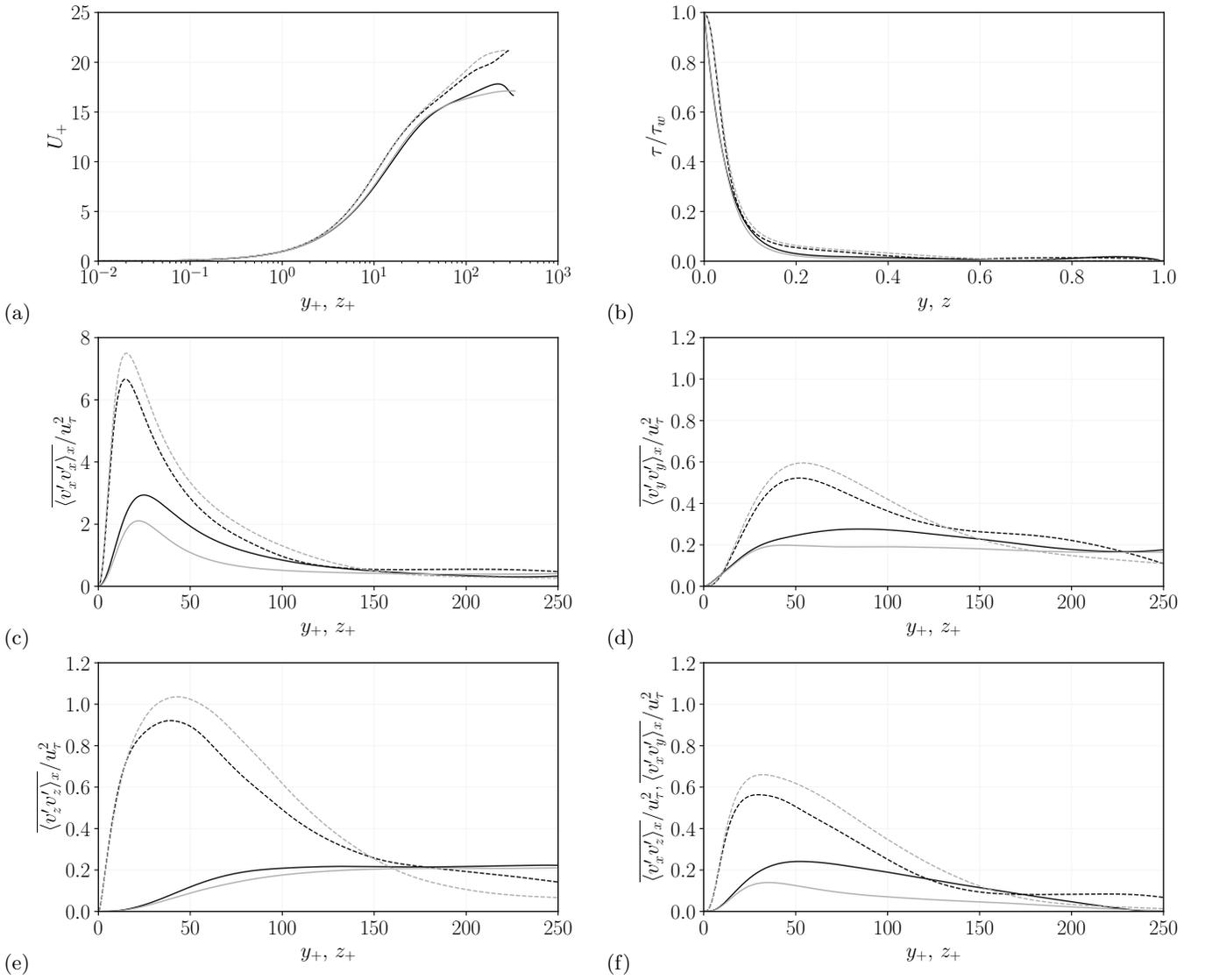

In Fig.~\ref{fig:turb_stat}, the profiles for the mean flow and the Reynolds stresses across the midlines of the quarter duct are compared with the corresponding (symmetrized) profiles of the full duct from Ref.~\cite{brynjell-rahkola_2024}.
Fig.~\ref{fig:stat_Up} contains profiles of 
$U_+^{\mathrm{H}}=\overline{\langle v_x\rangle_x}|_{y=0.5,z}/u_{\tau}^{\mathrm{H}}|_{y=0.5}$ and 
$U_+^{\mathrm{S}}=\overline{\langle v_x\rangle_x}|_{y,z=0.5}/u_{\tau}^{\mathrm{S}}|_{z=0.5}$.
It shows that the mean flow is almost unaffected by the symmetry condition, except in the log-law region $y_+,z_+ > 30$ \cite{pope2001turbulent} and beyond where the boundary conditions are felt and thus slight differences appear.
The normalized shear stress profiles
$\tau^{\mathrm{H}}|_{y=0.5,z}/\tau_w^{\mathrm{H}}|_{y=0.5}$ and
$\tau^{\mathrm{S}}|_{y,z=0.5}/\tau_w^{\mathrm{S}}|_{z=0.5}$ evaluated in the quarter and the full duct, are also close to indistinguishable (see Fig.~\ref{fig:stat_tau}).
Comparison of the Reynolds stresses \cite{pope2001turbulent} in Fig.~\ref{fig:stat_uu}--\ref{fig:stat_uw_uv} reveals that the turbulence intensities in the Shercliff layer are consistently underpredicted in the quarter duct compared to the full duct, whereas the corresponding quantities in the Hartmann layer are consistently overpredicted. 
These differences between the quarter and the full duct are visible outside the viscous sublayer $y_+,z_+ > 5$ \cite{pope2001turbulent}, but the stress profiles remain qualitatively similar.
(The Reynolds stress component $\overline{\langle v_y' v_z' \rangle}$ is small in comparison to the other stresses and is hence omitted in Fig.~\ref{fig:turb_stat}.)

% Bibliography
%apsrev4-2.bst 2019-01-14 (MD) hand-edited version of apsrev4-1.bst
%Control: key (0)
%Control: author (8) initials jnrlst
%Control: editor formatted (1) identically to author
%Control: production of article title (0) allowed
%Control: page (0) single
%Control: year (1) truncated
%Control: production of eprint (0) enabled
\providecommand{\noopsort}[1]{}

\end{document}